\documentclass[twocolumn,prb]{revtex4}
\usepackage{amsfonts}
\usepackage[T1]{fontenc}
\usepackage{amsmath,amsbsy,amssymb,graphicx}
\usepackage{times}

\begin{document}

\title{ Universal quantum gates, artificial neurons and pattern recognition\\
simulated by \textit{LC} resonators}
\author{Motohiko Ezawa}
\affiliation{Department of Applied Physics, University of Tokyo, Hongo 7-3-1, 113-8656,
Japan}

\begin{abstract}
We propose to simulate quantum gates by \textit{LC} resonators, where the
amplitude and the phase of the voltage describe the quantum state. By
controlling capacitance or inductance of resonators, it is possible to
control the phase of the voltage arbitrarily. A set of resonators acts as
the phase-shift, the Hadamard and the CNOT gates. They constitute a set of
universal quantum gates. We also discuss an application to an artificial
neuron. As an example, we study a pattern recognition of numbers and
alphabets by evaluating the similarity between an input and the reference
pattern. We also study a colored pattern recognition by using a complex
neural network.
\end{abstract}

\maketitle

\section{Introduction}

\label{SecIntro}

Quantum computation is one of the most exciting fields of current physics%
\cite{Feynman,DiVi,Nielsen}. It is realized in various systems including
superconductors\cite{Nakamura}, photonic systems\cite{Knill}, quantum dots%
\cite{Loss}, trapped ions\cite{Cirac} and nuclear magnetic resonance\cite%
{Vander,Kane}. For universal quantum computation, it is well known that only
three quantum gates are enough, which are the phase-shift, the Hadamard and
the CNOT gates\cite{Deutsch,Dawson,Universal}.

Recently, electric circuits attract renewed attention in the context of
topological physics\cite%
{ComPhys,TECNature,Garcia,Hel,Rosen,Lu,EzawaTEC,Hu,Hofmann,Research,EzawaLCR,EzawaNH,EzawaMajo,HelSkin}%
. There are also some attempts to simulate various quantum gates by electric
circuits\cite{EzawaUniv,EzawaDirac,EzawaTQC}. Among them, a network of
telegrapher lines is capable to simulate the universal quantum gates\cite%
{EzawaUniv,EzawaDirac}, because we may rewrite the Kirchhoff law in the form
of the Schr\"{o}dinger equation\cite{EzawaSch}. This formulation requires
long wires for a long quantum algorithm, where quantum states evolve
spatially from the left wires to the right wires.

Quantum machine learning is an emerging field of contemporary physics \cite%
{Lloyd,Schuld,Biamonte,Wittek,Harrow,Wiebe,Reben,ZLi,SchuldB,Hav,Lamata,Cong}%
. Neural networks are often used in machine learning, where artificial
neurons are basic elements\cite{Deep,Zurada}. An artificial neuron has an
internal degree of freedom called the weight. The output is determined from
the input data by taking the inner product between the input data and the
weight, and by\ applying an activation function to it. The inner product of
two objects measures the similarity between them. For instance, using a
series of numbers representing a reference pattern as the weight, we may
analyze the similarity between an input pattern and the reference pattern as
an output. Artificial neuron is simulated by quantum computer\cite%
{SchuldA,WiebeA,Cao,Tacc,Torro,Kris,Killo}. Taking the inner product is the
heaviest process, which will be executable by a quantum computer\cite%
{Tacc,Mangi}.

In this paper, we propose to simulate one qubit by a pair of \textit{LC}
resonators, where a set of voltage and current represents a wave function.
First, we construct a phase-shift gate with an arbitrary phase by tuning the
capacitance of an \textit{LC} resonator. Next, we construct the Hadamard
gate by tuning the inductance of an inductor bridging two \textit{LC}
resonators. We also construct the CNOT and the controlled phase-shift gate
by using a voltage-controlled inductor or capacitor. Finally, we discuss
applications to artificial neuron and pattern recognition. The calculation
of the inner product may be executed by the operation of \textit{LC}
resonators for arbitrary inputs and weights. We elucidate the difference
between the standard quantum-circuit implementation and the present
electric-circuit implementation of the inner product.

This paper is composed as follows. In Sec.\ref{SecLC}, we start with a
discussion how to store the information of $N$ qubits $\left\vert
j\right\rangle \!\rangle \equiv |n_{1}n_{2}\cdots n_{N}\rangle $ in a set of 
\textit{LC} resonators, where $n_{i}=0,1$. Then, we propose to construct
various quantum gates including a set of universal quantum gates by \textit{%
LC} resonators.

In Sec.\ref{SecArtificial}, we apply our formalism to study artificial
neurons, where we express various data with the aid of so-called real
equally weighted (REW) states\cite{DJ,Grover,HyperGraph}. They are
superposition states of $N$-qubits with coefficients $\alpha _{j}=\pm 1/%
\sqrt{2^{N}}$. In Sec.\ref{SecPattern}, we discuss a pattern recognition by
calculating the inner product of an input data and the reference data. We
present explicit examples of number recognition and alphabet recognition.

In Sec.\ref{SecComplex}, we generalize REW states to include complex
coefficients $\alpha _{j}=e^{i\theta _{j}}/\sqrt{2^{N}}$. We call them
complex equally weighted (CEW) states. Then, we introduce complex-artificial
neurons to deal with the inner product of CEW states. In Sec.\ref{SecColor},
we propose to represent a colored pattern by a CEW state, where colored
pattern recognition is done by evaluating the inner product of two CEW
states representing the reference and an input pattern.

In Sec.\ref{SecEC}, we present an electric-circuit implementation of quantum
gates for calculation of an inner product starting from the initial $N$%
-qubit state $|00\cdots 0\rangle $.

In Sec.\ref{SecGraph}, we explore REW states from a viewpoint of graph and
hypergraph states. We also introduce weighted graph and hypergraph states to
represent CEW states. Sec.\ref{SecDisc} is devoted to discussions.

\section{\textit{LC} resonators, qubits and gates}

\label{SecLC}

We use a set of $2^{N}$ identical \textit{LC} resonators to simulate $N$%
-qubit quantum computation. An instance of $N=2$ is illustrated in Fig.\ref%
{FigGate}(a). The voltage of the $j$th \textit{LC} resonator is expressed as%
\begin{equation}
V_{j}\left( t\right) =V_{j}^{0}\cos \left( \omega _{0}t+\theta _{j}\right) ,
\label{Vt}
\end{equation}%
where $\omega _{0}=1/\sqrt{LC}$ is the resonant frequency, $V_{j}^{0}$ is
the absolute value of the voltage and $\theta _{j}$ is the phase shift.

A qubit state is defined by a superposition of the two states $\left\vert
0\right\rangle $ and $\left\vert 1\right\rangle $ as $\left\vert \psi
\right\rangle =\alpha _{0}\left\vert 0\right\rangle +\alpha _{1}\left\vert
1\right\rangle $. Similarly, an $N$-qubit state is defined by a
superposition of the $2^{N}$ states as%
\begin{equation}
\left\vert \psi \right\rangle =\sum_{n_{j}=0,1}\alpha _{n_{1}n_{2}\cdots
n_{N}}|n_{1}n_{2}\cdots n_{N}\rangle ,
\end{equation}%
which is expressed equivalently as%
\begin{equation}
\left\vert \psi \right\rangle =\sum_{j=0}^{2^{N}-1}\alpha _{j}\left\vert
j\right\rangle \!\rangle ,  \label{N-qubit}
\end{equation}%
where $j$\ is the decimal number corresponding to the binary nuber $%
(n_{1}n_{2}\cdots n_{N})$ such as $\left\vert 0\right\rangle \!\rangle
=\left\vert 0\cdots 00\right\rangle $, $\left\vert 1\right\rangle \!\rangle
=\left\vert 0\cdots 01\right\rangle $, $\cdots $ , $\left\vert
2^{N}-1\right\rangle \!\rangle =\left\vert 11\cdots 1\right\rangle $.

It is a key observation\cite{EzawaTQC} that we may set 
\begin{equation}
\alpha _{j}=V_{j}^{0}e^{i\theta _{j}}/\sqrt{\sum_{j}(V_{j}^{0})^{2}}
\label{Aj}
\end{equation}%
in the \textit{LC}-resonator realization of quantum computation. Thus we
store the information of $N$ qubits in a set of \textit{LC} resonators.

\begin{figure}[t]
\centerline{\includegraphics[width=0.48\textwidth]{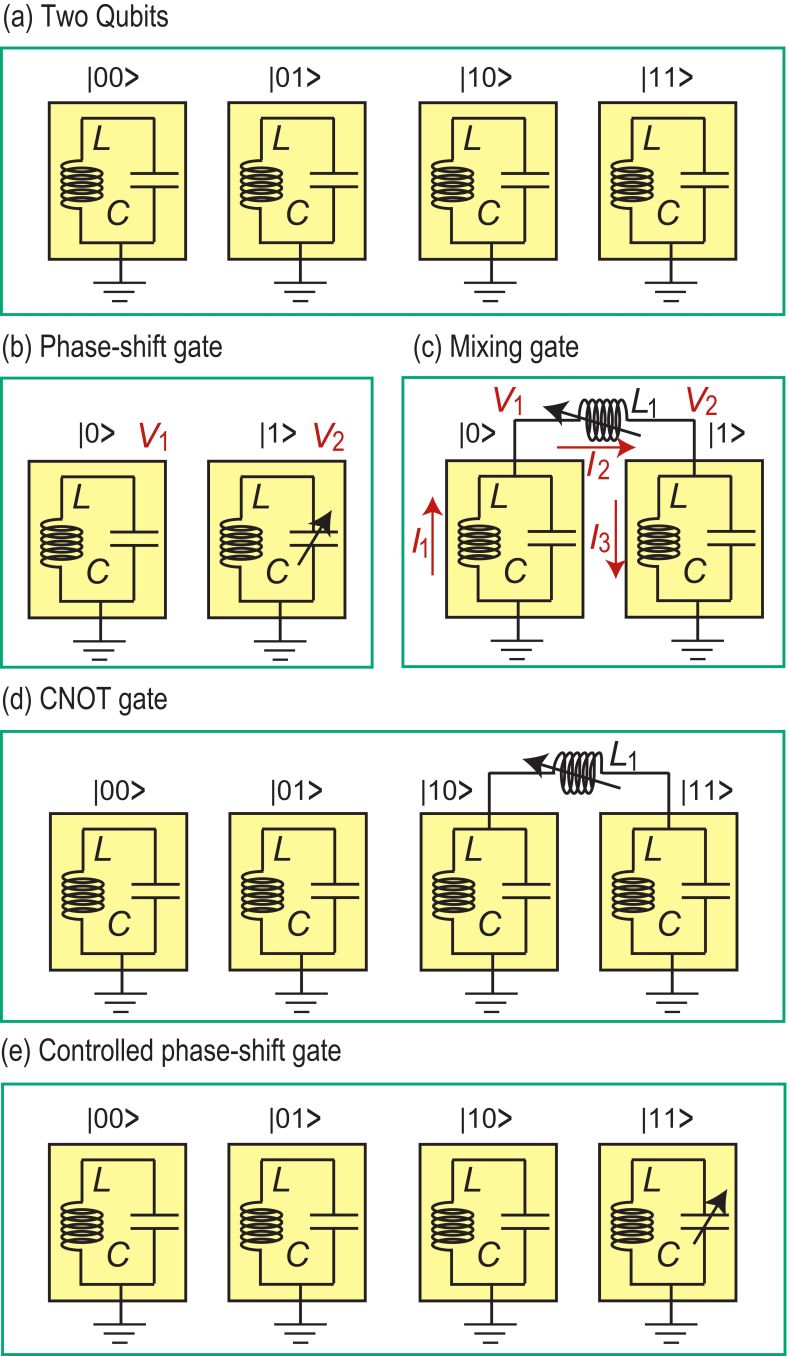}}
\caption{(a) Two qubits made of four \textit{LC} resonators. (b) Phase-shift
gate consisting a pair of \textit{LC} resonators. The capacitance of the
state $|1\rangle $ is controlled. (c) Mixing gate. The inductance of the
inductor $L_{1}$ bridging two resonators is controlled. The Hadamard gate is
constructed by a combination of the mixing gate and the phase-shift gate.
(d) CNOT gate. We bridge the resonators representing $|10\rangle $ and $%
|11\rangle $ by the inductor $L_{1}$ (e) Controlled phase-shift gate. We
control the capacitance of the resonator representing $|11\rangle $. }
\label{FigGate}
\end{figure}

Here we propose to carry out a gate process by controlling externally the
value $C$ of a capacitance as in Fig.\ref{FigGate}(b) or the value $L_{1}$
of an inductor bridging two \textit{LC} resonators as in Fig.\ref{FigGate}%
(c). For each gate process the initial and the final systems are the same
set of $2^{N}$ identical \textit{LC} resonators with the same energy,
although the coefficient $\alpha _{j}$ may be modified for some $j$. A gate
process is required to be adiabatic.

The gate $U$ is represented by a $2^{N}\times 2^{N}$ matrix $U_{jk}$ such
that%
\begin{equation}
U\left\vert j\right\rangle \!\rangle =\sum_{k=0}^{2^{N}-1}U_{jk}\left\vert
k\right\rangle \!\rangle .  \label{Gate}
\end{equation}%
By this operation, the initial state $\psi ^{\text{ini}}$ is brought to the
final state $\psi ^{\text{fin}}=U\psi ^{\text{ini}}$, where $\psi ^{\text{ini%
}}=\sum_{j=0}^{2^{N}-1}\alpha _{j}^{\text{ini}}\left\vert j\right\rangle
\!\rangle $ and $\psi ^{\text{fin}}=\sum_{k=0}^{2^{N}-1}\alpha _{k}^{\text{%
fin}}\left\vert k\right\rangle \!\rangle $. It follows that%
\begin{equation}
\alpha _{k}^{\text{fin}}=\sum_{j=0}^{2^{N}-1}\alpha _{j}^{\text{ini}%
}U_{jk}=\sum_{j=0}^{2^{N}-1}U_{kj}\alpha _{j}^{\text{ini}},  \label{GateU}
\end{equation}%
since $U$ is a symmetric matrix in universal quantum computation.

\textbf{Kirchhoff law and Schr\"{o}dinger equation.} We first consider a set
of independent \textit{LC} resonators. The Kirchhoff law of the $j$th 
\textit{LC} resonator may be rewritten in the form of the Schr\"{o}dinger
equation\cite{EzawaSch,EzawaUniv},%
\begin{equation}
i\frac{d}{dt}\psi _{j}=H\psi _{j},  \label{Sch}
\end{equation}%
where $H\left( t\right) =-\omega _{0}\sigma _{y}$ is the Hamiltonian, and 
\begin{equation}
\psi _{j}=\left( \mathcal{I}_{j},\mathcal{V}_{j}\right) ^{t}=\left( \sqrt{L/C%
}I_{j},V_{j}\right) ^{t}  \label{PsiJ}
\end{equation}%
is the wave function.

\textbf{Energy conservation and probability conservation.} The total energy
of the system is given by $U_{\text{T}}=U_{\text{E}}+U_{\text{M}}$ with%
\begin{equation}
U_{\text{E}}=\frac{C}{2}\sum_{j}V_{j}^{2},\qquad U_{\text{M}}=\frac{L}{2}%
\sum_{j}I_{j}^{2},
\end{equation}%
where $U_{\text{E}}$ and $U_{\text{M}}$\ are the electrostatic energy and
the magnetic energy, respectively.

On the other hand, by using (\ref{PsiJ}), the probability of the wave
function is rewritten as 
\begin{equation}
\sum_{j}\left\vert \psi _{j}\right\vert ^{2}=\sum_{j}\mathcal{I}_{j}^{2}+%
\mathcal{V}_{j}^{2}=\sum_{j}\frac{L}{C}I_{j}^{2}+V_{j}^{2}=\frac{2}{C}U_{%
\text{T}}.
\end{equation}%
Hence, the conservation of the probability of the wave function is assured
by the conservation of the total energy\cite{EzawaDirac}. As we have stated,
we arrange a gate process so that the total energy is the same before and
after the gate process. It corresponds to the conservation of the
probability for qubits\textbf{\ }$\sum_{j}\left\vert \alpha _{j}\right\vert
^{2}=1$.

\begin{figure}[t]
\centerline{\includegraphics[width=0.48\textwidth]{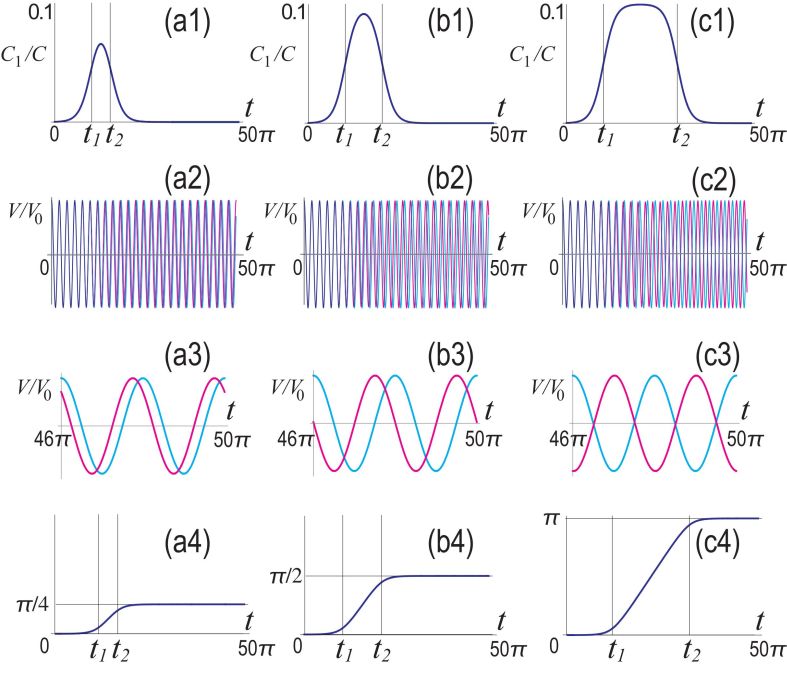}}
\caption{Phase-shift gate for (a) $\protect\phi =\protect\pi /4$, (b) $%
\protect\phi =\protect\pi /2$ and (c) $\protect\phi =\protect\pi $. The
phase delay is controlled by a time duration of the capacitance perturbation 
$C_{1}\left( t\right) $. (*1) Time evolution of the perturbed capacitance.
(*2) Time evolution of the voltage $V_{2}$. The voltage $V_{2}$ with
(without) the perturbation is represented by a magenta (cyan) curve. (*3)
Voltage $V_{2}$ in the final states. (*4) Phase delay as a function of time $%
t$. The horizontal axis is time $t$. Time span is $0<t<50\protect\pi $ for
(*1), (*2) and *(4). It is $46\protect\pi <t<50\protect\pi $ for (*3),
representing the final state. We set $C_{1}^{0}/C_{0}=0.1$ and $T=10\protect%
\omega _{0}$ for $0<\protect\omega _{0}t<50\protect\pi $. We set $\protect%
\omega _{0}t_{1}=10\protect\pi $ for all phase-shift gate. We also set $%
\protect\omega _{0}t_{2}=15\protect\pi $ for the $\protect\pi /4$
phase-shift gate, $\protect\omega _{0}t_{2}=20\protect\pi $ for the $\protect%
\pi /2$ phase-shift gate and $\protect\omega _{0}t_{2}=30\protect\pi $ for
the $\protect\pi $ phase-shift (i.e., Pauli Z) gate, }
\label{FigPhase}
\end{figure}

\textbf{Phase-shift gate.} The phase-shift gate is defined by the matrix%
\begin{equation}
U_{\phi }=\left( 
\begin{array}{cc}
1 & 0 \\ 
0 & e^{i\phi }%
\end{array}%
\right) ,
\end{equation}%
which acts on the one-qubit state $\left( \left\vert 0\right\rangle
,\left\vert 1\right\rangle \right) ^{t}$. Namely, the action is 
\begin{equation}
U_{\phi }\left\vert 0\right\rangle =\left\vert 0\right\rangle ,\qquad
U_{\phi }\left\vert 1\right\rangle =e^{i\phi }\left\vert 1\right\rangle .
\end{equation}%
To generate the phase shift $\phi $ in the wave function, it is enough to
control only the capacitance $C$ in the \textit{LC} resonator externally
during the gating process as shown in Fig.\ref{FigGate}(b). We control the
capacitance as $C\left( t\right) =C_{0}+C_{1}\left( t\right) $, where%
\begin{equation}
C_{1}\left( t\right) =\frac{C_{1}^{0}}{2}\left( \tanh \frac{t-t_{1}}{T}%
-\tanh \frac{t-t_{2}}{T}\right) ,  \label{FuncC1}
\end{equation}%
with four parameters $C_{1}^{0}$, $t_{1}$, $t_{2}$ and $T$, as shown in Fig.%
\ref{FigPhase}(a1), (b1) and (c1).

It is possible to determine analytically how the phase shift $\phi $ depends
on these parameters by calculating the Berry phase. Since the voltage
evolution $V\left( t\right) $ is written as a Schr\"{o}dinger equation, we
may use an adiabatic approximation. The snap shot wave function at time $%
t=\tau $ is given by $\psi (\tau )=\sqrt{2L/C}I_{0}\bar{\psi}(\tau )$, where 
$\bar{\psi}(\tau )$ is the normalized wave function,%
\begin{equation}
\bar{\psi}(\tau )=\frac{1}{\sqrt{2}}\exp (i\omega _{\tau }\tau )\left( 
\begin{array}{c}
1 \\ 
-i%
\end{array}%
\right) ,
\end{equation}%
with $\omega _{\tau }$ the snapshot frequency,%
\begin{equation}
\omega _{\tau }=1/\sqrt{LC\left( \tau \right) }.
\end{equation}%
The Berry phase is calculated as 
\begin{align}
\gamma =& i\int_{0}^{t}\left\langle \bar{\psi}\left( \tau \right)
\right\vert \partial _{\tau }\left\vert \bar{\psi}\left( \tau \right)
\right\rangle d\tau  \notag \\
=& \int_{0}^{t}\frac{1}{2\sqrt{L}C\left( \tau \right) ^{3/2}}\left( 2C\left(
\tau \right) -\tau \frac{dC\left( \tau \right) }{d\tau }\right) d\tau  \notag
\\
=& \int_{0}^{t}\left[ \omega \left( \tau \right) -\frac{\tau }{\sqrt{L}%
C\left( \tau \right) ^{3/2}}\frac{dC\left( \tau \right) }{d\tau }\right]
d\tau .
\end{align}%
When the perturbation $C_{1}\left( t\right) $ is small enough with respect
to $C_{0}$, it is calculated as%
\begin{equation}
\gamma \simeq i\omega _{0}t-\phi ,
\end{equation}%
where $\phi $ is the phase shift given by%
\begin{equation}
\phi =\omega _{0}\int_{0}^{t}\frac{C_{1}\left( \tau \right) }{2C}d\tau .
\end{equation}%
It is explicitly calculated as%
\begin{align}
\phi =& \omega _{0}\frac{C_{1}^{0}}{2C}T\left( \log \cosh \frac{t-t_{1}}{T}%
-\log \cosh \frac{t_{1}}{T}\right.  \notag \\
& \qquad \quad \left. -\log \cosh \frac{t-t_{2}}{T}+\log \cosh \frac{t_{2}}{T%
}\right) ,
\end{align}%
which yields%
\begin{equation}
\phi =\frac{C_{1}^{0}\omega _{0}}{2C}\left( t_{2}-t_{1}\right) ,
\label{PhaseShift}
\end{equation}%
provided $T\ll t_{1}<t_{2}\ll t$. Hence, we can tune the phase shift $\phi $
arbitrary by controlling the magnitude of $\left( t_{2}-t_{1}\right)
C_{1}^{0}$.

We next solve numerically the differential equation (\ref{Sch}) to study the
time evolution of the voltage $V(t)$, and confirm the phase-shift formula (%
\ref{PhaseShift}). When we fix $C_{1}^{0}=0.1C_{0}$\ and $\omega
_{0}t_{1}=10\pi $, we have%
\begin{equation}
\phi =\frac{1}{20}\left( \omega _{0}t_{2}-10\pi \right) .
\end{equation}%
We present numerical results of the time evolution $V(t)$ by choosing $%
\omega _{0}t_{2}=15\pi $, $20\pi $ and $30\pi $ in Fig.\ref{FigPhase}(a2),
(b2) and (c2). See Fig.\ref{FigPhase}(a3), (b3) and (c3) for $V(t)$ for $%
t\gg t_{2}$, representing the final state. The phase shift is found to occur
due to the perturbation $C_{1}\left( t\right) $. The phase shift $\phi (t)$
during a gating process is shown in Fig.\ref{FigPhase}(a4), (b4) and (c4).
After the gating process, the resonance frequency returns to $\omega _{0}$
but the phase $\phi $ becomes different from the initial value. It reads $%
\phi =\pi /4,\pi /2$\ and $\pi $ as in Fig.\ref{FigPhase}(a4), (b4) and
(c4). These numerical results confirm the analytical formula (\ref%
{PhaseShift}).

\textbf{Hadamard gate.} The Hadamard gate is defined by the matrix%
\begin{equation}
U_{\text{H}}\equiv \frac{1}{\sqrt{2}}\left( 
\begin{array}{cc}
1 & 1 \\ 
1 & -1%
\end{array}%
\right) ,
\end{equation}%
which acts the one-qubit state $\left( \left\vert 0\right\rangle ,\left\vert
1\right\rangle \right) ^{t}$. It is known to be given by\cite%
{EzawaUniv,EzawaDirac}%
\begin{equation}
U_{\text{H}}=e^{-i\pi /4}U_{\pi /2}U_{\text{mix}}U_{\pi /2},
\end{equation}%
where $U_{\pi /2}$ is the $\pi /2$ phase-shift gate, while $U_{\text{mix}}$
is the mixing gate defined by%
\begin{equation}
U_{\text{mix}}=\frac{1}{\sqrt{2}}\left( 
\begin{array}{cc}
e^{i\pi /4} & e^{-i\pi /4} \\ 
e^{-i\pi /4} & e^{i\pi /4}%
\end{array}%
\right) .  \label{Mix}
\end{equation}%
We construct the mixing gate (\ref{Mix}) in what follows.

We consider a pair of \textit{LC} resonators bridged by an inductor $L_{1}$
as shown in Fig.\ref{FigGate}(c), where the inductance $L_{1}$\ is
controlled externally. The Kirchhoff law reads%
\begin{equation}
\frac{d}{dt}\left( 
\begin{array}{c}
I_{1} \\ 
I_{2} \\ 
I_{3} \\ 
V_{1} \\ 
V_{2}%
\end{array}%
\right) =\left( 
\begin{array}{ccccc}
0 & 0 & 0 & \frac{1}{L} & 0 \\ 
0 & 0 & 0 & -\frac{1}{L_{1}} & \frac{1}{L_{1}} \\ 
0 & 0 & 0 & 0 & -\frac{1}{L} \\ 
-\frac{1}{C} & \frac{1}{C} & 0 & 0 & 0 \\ 
0 & -\frac{1}{C} & \frac{1}{C} & 0 & 0%
\end{array}%
\right) \left( 
\begin{array}{c}
I_{1} \\ 
I_{2} \\ 
I_{3} \\ 
V_{1} \\ 
V_{2}%
\end{array}%
\right) ,  \label{Kirchhoff}
\end{equation}%
where $I_{1}$, $I_{2}$, $I_{3}$, $V_{1}$ and $V_{2}$ are defined in Fig.\ref%
{FigGate}(c). It is rewritten in the form of the Schr\"{o}dinger equation as
in Eq.(\ref{Sch}) with the Hamiltonian 
\begin{equation}
H=\frac{1}{\sqrt{LC}}\left( 
\begin{array}{ccccc}
0 & 0 & 0 & i & 0 \\ 
0 & 0 & 0 & -iL/L_{1}\left( t\right) & iL/L_{1}\left( t\right) \\ 
0 & 0 & 0 & 0 & -i \\ 
-i & i & 0 & 0 & 0 \\ 
0 & -i & i & 0 & 0%
\end{array}%
\right) ,
\end{equation}%
and the wave function 
\begin{equation}
\left( \mathcal{I}_{1},\mathcal{I}_{2},\mathcal{I}_{3},\mathcal{V}_{1},%
\mathcal{V}_{2}\right) =\left( \sqrt{\frac{L}{C}}I_{1},\sqrt{\frac{L}{C}}%
I_{2},\sqrt{\frac{L}{C}}I_{3},V_{1},V_{2}\right) .
\end{equation}%
By making a snapshot approximation, the eigenvalues are given by%
\begin{equation}
E=0,\pm \omega _{0},\pm \ell \left( t\right) \omega _{0},
\end{equation}%
with%
\begin{equation}
\ell \left( t\right) =\sqrt{1+2L/L_{1}\left( t\right) }
\end{equation}%
at each $t$.

\begin{figure}[t]
\centerline{\includegraphics[width=0.48\textwidth]{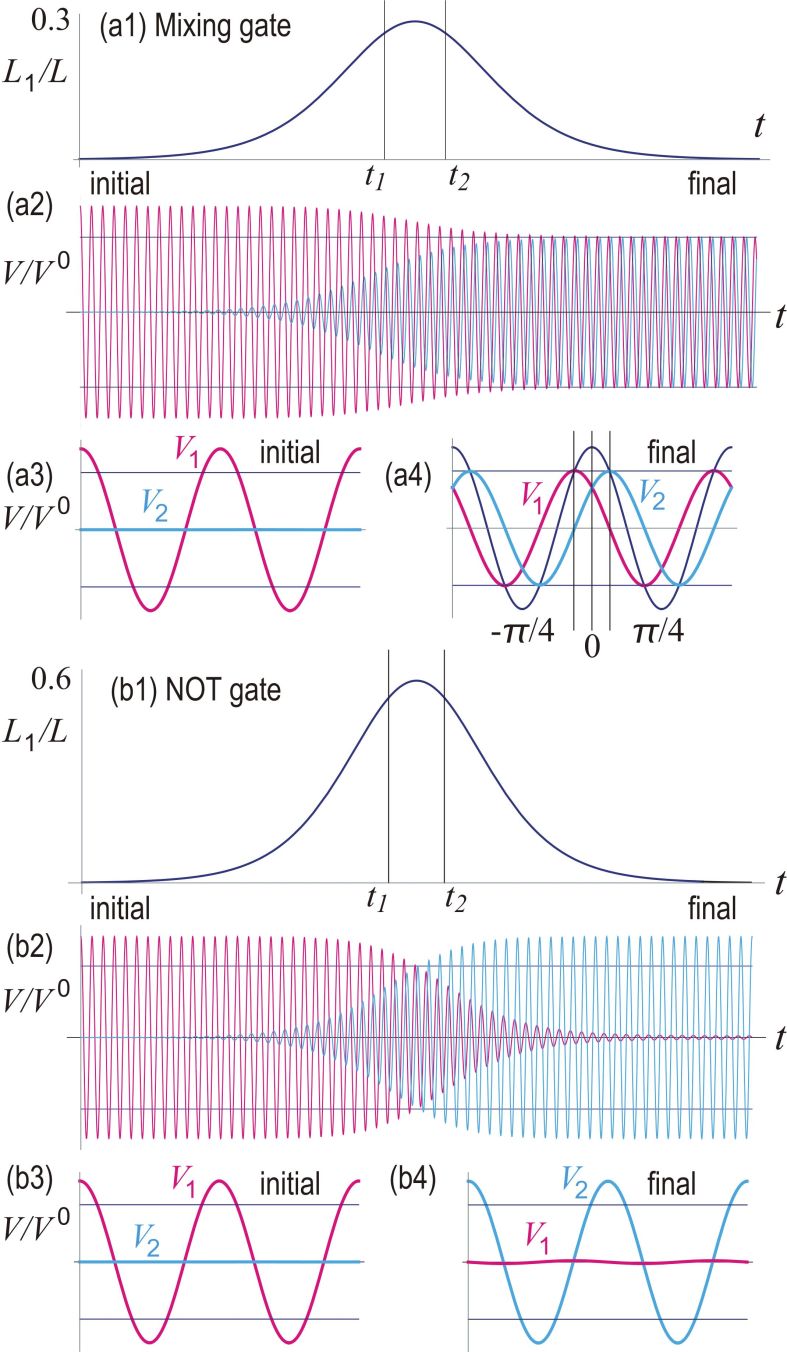}}
\caption{(a) Mixing gate. (b) NOT gate. (a1) and (b1) Time-depending
perturbation introduced to the inductance. (a2) and (b2) Time evolution of
the voltage $V_{1}$ ($V_{2}$) at the left (right) \textit{LC} resonator is
represented by a magenta (cyan) curve. (*3) Time evolution of $V_{1}$ and $%
V_{2}$ for $0<\protect\omega _{0}t<4\protect\pi $, represending the initial
state, where $V_{2}=0$. (*4) Time evolution $V_{1}$ and $V_{2}$ for $116%
\protect\pi <t<120\protect\pi $, representing the final state, where the
voltage without the perturbation is represented by a black curve. We set $%
L_{1}/L=0.1$ and $T=10\protect\omega _{0}$ for $0<\protect\omega _{0}t<120%
\protect\pi $. We set $\protect\omega _{0}t_{1}=50\protect\pi $ and $\protect%
\omega _{0}t_{2}=55\protect\pi $ for the mixing gate and $\protect\omega %
_{0}t_{1}=55\protect\pi $ and $\protect\omega _{0}t_{2}=65\protect\pi $ for
the NOT gate. The orange lines represent the voltage $\pm V_{0}/\protect%
\sqrt{2}$. }
\label{FigHadamard}
\end{figure}

We consider a process where the inductor $L_{1}$ is bridged to the \textit{LC%
} resonators during a time interval $t_{1}<t<t_{2}$ but not for $t<t_{1}$
and $t>t_{2}$. For example, we may take%
\begin{equation}
\frac{1}{L_{1}\left( t\right) }=\frac{1}{2L_{1}}\left( \tanh \frac{t-t_{1}}{T%
}-\tanh \frac{t-t_{2}}{T}\right) ,  \label{L1}
\end{equation}%
which we have illustrated in Fig.\ref{FigHadamard}(a).

We solve (\ref{Kirchhoff}) numerically with the use of (\ref{L1}) and show
how the voltage evolves in Fig.\ref{FigHadamard}. By tuning $t_{2}-t_{1}$
and $L_{1}$ appropriately, in order to construct the mixing gate (\ref{Mix}%
), we make the magnitudes of $V_{1}$ and $V_{2}$ identical in the final
state, i.e., for $t\gg t_{2}$. We find the phase delay $\pi /4$ in $V_{1}$
and the phase advance $\pi /4$ in $V_{2}$ as in Fig.\ref{FigHadamard}(a4).

We may discuss the process analytically. For this purpose, we approximate (%
\ref{L1}) by a step function such that $1/L_{1}\left( t\right) =0$ for $%
t<t_{1}$, $L_{1}\left( t\right) =L_{1}$ for $t_{1}<t<t_{2}$ and $%
1/L_{1}\left( t\right) =0$ for $t>t_{2}$. Two resonators are decoupled when $%
1/L_{1}\left( t\right) =0$. For definiteness we choose $t_{1}=0$.

First, we analyze the case where only the left \textit{AC} resonator is
active for $t\leq 0$, or%
\begin{equation}
V_{1}^{\text{ini}}\left( t\right) =V_{0}\cos (\omega _{0}t),\qquad V_{2}^{%
\text{ini}}\left( 0\right) =0.  \label{V0}
\end{equation}%
At $t=t_{1}$, the perturbation $L_{1}(t)$ is set on.

(i) For $0\leq t\leq t_{2}$, we may solve the Kirchhoff equation (\ref%
{Kirchhoff}) for the voltages as%
\begin{align}
V_{1}\left( t\right) =& V_{0}\cos \left[ \frac{\ell +1}{2}\omega _{0}t\right]
\cos \left[ \frac{\ell -1}{2}\omega _{0}t\right] ,  \label{V1} \\
V_{2}\left( t\right) =& V_{0}\sin \left[ \frac{\ell +1}{2}\omega _{0}t\right]
\sin \left[ \frac{\ell -1}{2}\omega _{0}t\right] ,  \label{V2}
\end{align}%
where we have chosen the initial condition to meet (\ref{V0}), or%
\begin{equation}
V_{1}\left( 0\right) =V_{0},\qquad V_{2}\left( 0\right) =0.
\end{equation}%
When $\ell \simeq 1$, the oscillation modes are made of the high-frequency
mode $\frac{\ell +1}{2}\omega _{0}$ and the low-frequency mode $\frac{\ell -1%
}{2}\omega _{0}$.

\begin{figure*}[t]
\centerline{\includegraphics[width=0.88\textwidth]{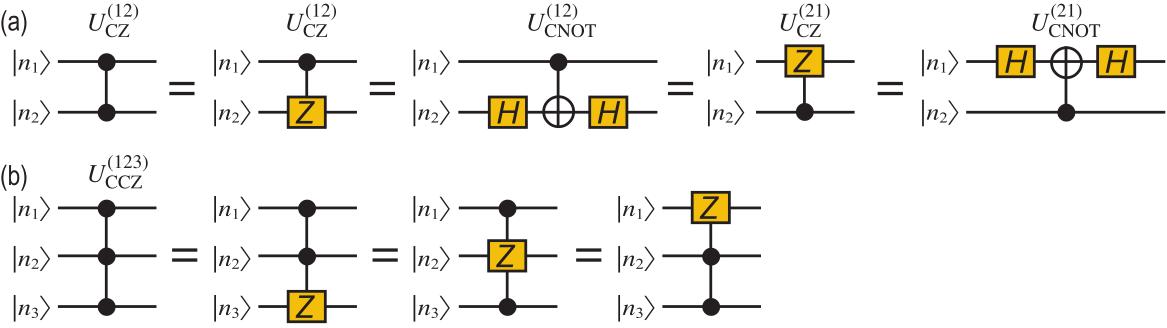}}
\caption{Equivalent quantum-circuit representation of (a) CZ gate and (b)
CCZ gate.}
\label{FigCZ}
\end{figure*}

(ii) At $t=t_{2}$, we require the amplitudes of $V_{1}\left( t\right) $ and $%
V_{2}\left( t\right) $ to be identical. Since the amplitude is determined by
the low-frequency mode, the condition reads 
\begin{equation}
\cos \left[ \frac{\ell -1}{2}\omega _{0}t_{2}\right] =\sin \left[ \frac{\ell
-1}{2}\omega _{0}t_{2}\right] =\frac{1}{\sqrt{2}}.
\end{equation}%
Since the connection is weak, we have $L/L_{1}\ll 1$, which leads to $\ell
\simeq 1+L/L_{1}$. We use it to derive the relation%
\begin{equation}
\frac{L}{2L_{1}}\omega _{0}t_{2}=\frac{\pi }{4},  \label{EqA}
\end{equation}%
which fixes $t_{2}$ to generate the mixing gate (\ref{Mix}). The voltages
read%
\begin{align}
V_{1}\left( t_{2}\right) =& \frac{V_{0}}{\sqrt{2}}\cos \left[ \frac{\ell +1}{%
2}\omega _{0}t_{2}\right] =\frac{V_{0}}{\sqrt{2}}\cos \left[ \omega
_{0}t_{2}+\frac{\pi }{4}\right] ,  \label{V3} \\
V_{2}\left( t_{2}\right) =& \frac{V_{0}}{\sqrt{2}}\sin \left[ \frac{\ell +1}{%
2}\omega _{0}t_{2}\right] =\frac{V_{0}}{\sqrt{2}}\cos \left[ \omega
_{0}t_{2}-\frac{\pi }{4}\right] ,  \label{V4}
\end{align}%
where use was made of (\ref{V1}), (\ref{V2}) and (\ref{EqA}). There are
phase shifts $\pm \frac{\pi }{4}$.

(iii) For $t>t_{2}$, since the perturbation is off, two \textit{LC}
resonators resonate independently with the initial condition (\ref{V3}) and (%
\ref{V4}), or%
\begin{align}
V_{1}^{\text{fin}}(t)=& \frac{V_{0}}{\sqrt{2}}\cos \left[ \omega _{0}t+\frac{%
\pi }{4}\right] ,  \label{V5} \\
V_{2}^{\text{fin}}(t)=& \frac{V_{0}}{\sqrt{2}}\cos \left[ \omega _{0}t-\frac{%
\pi }{4}\right] .  \label{V6}
\end{align}%
It followed that 
\begin{align}
\alpha _{1}^{\text{ini}}=& 1,\qquad \alpha _{2}^{\text{ini}}=0,  \label{A1}
\\
\alpha _{1}^{\text{fin}}=& \frac{1}{\sqrt{2}}e^{i\pi /4},\qquad \alpha _{2}^{%
\text{fin}}=\frac{1}{\sqrt{2}}e^{-i\pi /4}  \label{A2}
\end{align}%
from Eqs.(\ref{Aj}), (\ref{V0}), (\ref{V5}) and (\ref{V6}).

Next, we analyze the case where only the right AC resonator is active for $%
t\leq 0$, or%
\begin{equation}
V_{1}^{\text{ini}}\left( t\right) =0,\qquad V_{2}^{\text{ini}}\left(
0\right) =V_{0}\cos (\omega _{0}t)
\end{equation}%
instead of (\ref{V0}). By making precisely the same analysis, we obtain 
\begin{align}
\alpha _{1}^{\text{ini}}=& 0,\qquad \alpha _{2}^{\text{ini}}=1,  \label{A3}
\\
\alpha _{1}^{\text{fin}}=& \frac{1}{\sqrt{2}}e^{-i\pi /4},\qquad \alpha
_{2}^{\text{fin}}=\frac{1}{\sqrt{2}}e^{i\pi /4}.  \label{A4}
\end{align}%
The results (\ref{A1}), (\ref{A2}) (\ref{A3}) and (\ref{A4}) are summarized
as the mixing gate (\ref{Mix}) based on the definition (\ref{GateU}).

\textbf{NOT gate.} The NOT gate is defined by the matrix%
\begin{equation}
U_{\text{NOT}}=\left( 
\begin{array}{cc}
0 & 1 \\ 
1 & 0%
\end{array}%
\right) ,  \label{NOT}
\end{equation}%
which acts on one qubit. We find from Eq.(\ref{Mix}) that%
\begin{equation}
U_{\text{NOT}}=U_{\text{mix}}^{2}.
\end{equation}%
It is given by the sequential applications of the mixing gate. The
construction is similar to that of the mixing gate provided the duration of
the inductor $L_{1}$ is made twice. We present numerical results in Fig.\ref%
{FigHadamard}(b). With respect to an analytical study, the main equation is 
\begin{equation}
\frac{L}{2L_{1}}\omega _{0}t_{2}=\frac{\pi }{2}
\end{equation}%
in place of Eq.(\ref{EqA}).

\textbf{One qubit universal gate.} We may construct a combination of the
Hadamard and phase-shift gates such as%
\begin{equation}
U_{\text{1bit}}=e^{-i\theta /2}U_{\phi +\pi }U_{\text{H}}U_{\theta }U_{\text{%
H}}=\left( 
\begin{array}{cc}
\cos \frac{\theta }{2} & -i\sin \frac{\theta }{2} \\ 
ie^{i\phi }\sin \frac{\theta }{2} & -e^{i\phi }\cos \frac{\theta }{2}%
\end{array}%
\right) ,
\end{equation}%
which represents any SU(2) generator. It is called the one-qubit
universal-quantum gate.

\textbf{CNOT gate.} The CNOT gate is defined by a matrix%
\begin{equation}
U_{\text{CNOT}}=\left( 
\begin{array}{cccc}
1 & 0 & 0 & 0 \\ 
0 & 1 & 0 & 0 \\ 
0 & 0 & 0 & 1 \\ 
0 & 0 & 1 & 0%
\end{array}%
\right) ,
\end{equation}%
which acts the two-qubit state $\left( \left\vert 00\right\rangle
,\left\vert 01\right\rangle ,\left\vert 10\right\rangle ,\left\vert
11\right\rangle \right) ^{t}$. Two-qubit operation is constructed by using
four \textit{LC} resonators as in Fig.\ref{FigGate}. The CNOT gate is
constructed by applying the NOT gate between the resonators representing $%
\left\vert 10\right\rangle $ and $\left\vert 11\right\rangle $, as shown in
Fig.\ref{FigGate}(d).

\textbf{Controlled Z gate.} The CZ gate is defined by a matrix%
\begin{equation}
U_{\text{CZ}}=\text{diag.}\left[ 1,1,1,-1\right] ,  \label{CZgate}
\end{equation}%
which acts on the two-qubit state $\left( \left\vert 00\right\rangle
,\left\vert 01\right\rangle ,\left\vert 10\right\rangle ,\left\vert
11\right\rangle \right) ^{t}$. It follows from the definition that the
controlled and target qubits are symmetric in the CZ gate, which leads to
various equivalence quantum circuits as shown in Fig.\ref{FigCZ}(a). We
denote the CZ gate by the two black disks connected by a line.

\textbf{CCZ gate.} In a similar way to the CZ gate, we can construct the
controlled-controlled Z (CCZ) gate acting on three qubits. It flips the sign 
$x_{7}$ of the state $\left\vert 111\right\rangle $. Namely, we flip $x_{7}$
to $-x_{7}$. As in the case of the CZ gate, the CCZ gate is symmetric with
respect to the exchange of the controlled and target qubits shown in Fig.\ref%
{FigCZ}(b). We denote the CCZ gate by the three black disks connected by a
line.

\textbf{C}$^{p-1}$\textbf{Z gate.} We further generalize the CCZ gate to the
C$^{p-1}$Z gate. It is a $p$-qubit gate, which flips the sign $x_{2^{p}-1}$\
of the coefficient of the state $\left\vert 11\cdots 1\right\rangle $. As in
the case of the CZ and CCZ gates, the C$^{p-1}$Z gate is symmetric with
respect to the controlled and target qubits.

More generally, we may take an $N$-qubit system with $N>p$. We may consider
a C$^{p-1}$Z gate acting a $p$-qubit subspace. We denote it by $p$\ black
disks connected by a line. Such C$^{p-1}$Z gates play an essential role to
make a hypergraph state as we will soon see.

\textbf{Controlled phase-shift gate.} The controlled phase-shift gate is
defined by the matrix%
\begin{equation}
U_{\text{Z}_{\phi }}=\text{diag.}\left[ 1,1,1,e^{i\phi }\right] ,
\label{CPhi}
\end{equation}%
which acts on two qubits. There is no action on the target qubit if the
control qubit is $\left\vert 0\right\rangle $, while the $\phi $ phase-shift
gate is applied if the control qubit is $\left\vert 1\right\rangle $. The
controlled phase-shift gate is constructed by applying the phase-shift gate
for the \textit{LC} resonators representing $\left\vert 11\right\rangle $,
as shown in Fig.\ref{FigGate}(e).

Note that the CZ gate (\ref{CZgate}) is obtained by setting $\phi =\pi $ in
the controlled phase-shift gate (\ref{CPhi}). Namely, it may be viewed as a
generalization of the CZ gate, and hence we call it the CZ$_{\phi }$ gate.

\textbf{C}$^{p-1}$\textbf{Z}$_{\phi }$\textbf{\ gate.} In a similar way to C$%
^{p-1}$Z gates, we may define multi-controlled phase-shift gates, which we
denote by C$^{p-1}$Z$_{\phi }$ gates. It is a $p$-qubit gate, which
multiplies the phase $e^{i\phi }$\ to the coefficient $\alpha _{2^{p}-1}$ of
the state $\left\vert 11\cdots 1\right\rangle $.

\begin{figure}[t]
\centerline{\includegraphics[width=0.49\textwidth]{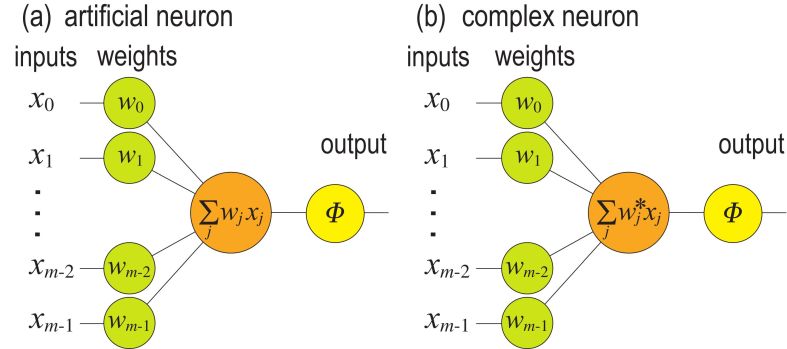}}
\caption{(a) Schematic for an artificial neuron. It has $m$ real inputs $%
x_{j}$ and real internal degree of freedom named weights $w_{j}$. Output is
obtained by calculating the inner product $\sum_{j}w_{j}x_{j}$ and then by
applying an activation function $\Phi (\sum_{j}w_{j}x_{j})$. (b) Schematic
for a complex neuron. It has complex $m$ inputs $x_{j}$ and complex weights $%
w_{j}$ with the inner product $\sum_{j}w^*_{j}x_{j}$.}
\label{FigNeuron}
\end{figure}

\section{Artificial neuron}

\label{SecArtificial}

An artificial neuron is a mathematical model\cite{Deep,Zurada} to simulate a
biological neuron. There are $m$ inputs $x_{0}$, $x_{1}$, $\cdots $, $%
x_{m-1} $ and $m$ weights $w_{0}$, $w_{1}$, $\cdots $, $w_{m-1}$, where, $%
x_{j}$ and $w_{j}$ are real numbers. We represent the input and the weight
by wave functions as\cite{Tacc} 
\begin{equation}
\left\vert \psi _{x}\right\rangle =\frac{1}{\sqrt{2^{N}}}%
\sum_{j=0}^{2^{N}-1}x_{j}\left\vert j\right\rangle \!\rangle ,\quad
\left\vert \psi _{w}\right\rangle =\frac{1}{\sqrt{2^{N}}}%
\sum_{j=0}^{2^{N}-1}w_{j}\left\vert j\right\rangle \!\rangle ,  \label{WaveA}
\end{equation}%
where $\left\vert j\right\rangle \!\rangle $ forms the $N$ qubit basis as in
Eq.(\ref{N-qubit}). Note the difference between the coefficients $\alpha
_{j} $ in Eq.(\ref{N-qubit}) and $x_{j}$, $w_{j}$ in Eq.(\ref{WaveA}) by the
factor $1/\sqrt{2^{N}}$.

The first step in the artificial neuron is to calculate the inner product $%
\sum_{j}w_{j}x_{j}$. The inner product of the input data and the weight data
measures the similarity between them. For instance, using a series of
numbers representing a set of reference patterns as the weight, we may
calculate the similarity between an input pattern and the reference pattern.

The inner product is outputted after applying an activation function,%
\begin{equation}
y=\Phi (\sum_{j}w_{j}x_{j}).  \label{ActiFunc}
\end{equation}%
The activation function $\Phi $ has various forms such as the step function%
\cite{RosenF,McC}, a linear function, a sigmoid function, a ramp function%
\cite{Glo} and so on. We show a schematic of a neuron in Fig.\ref{FigNeuron}%
(a). In the process of artificial neuron, the heaviest procedure is the
calculation of $\sum_{j}w_{j}x_{j}$, which is efficiently done by using a
quantum computer\cite{Tacc}.

We implement the wave functions (\ref{WaveA}) by unitary transformations
from the initial state $\left\vert 0\right\rangle \!\rangle $,%
\begin{equation}
\left\vert \psi _{x}\right\rangle =U_{x}\left\vert 0\right\rangle \!\rangle
,\qquad \left\vert \psi _{w}\right\rangle =U_{w}^{\dag }\left\vert
0\right\rangle \!\rangle .  \label{WaveB}
\end{equation}%
Then, the inner product is calculated as%
\begin{equation}
\sum_{j}w_{j}x_{j}=2^{N}\langle \psi _{w}|\psi _{x}\rangle =2^{N}\langle
\!\langle 0|U_{w}U_{x}\left\vert 0\right\rangle \!\rangle .
\label{InnerProd}
\end{equation}%
We explicitly construct $U_{x}$ and $U_{w}$ later in this section. On the
other hand, the application of $\Phi $ is easy with the use of a classical
computer since it is a one-to-one map.

\begin{figure}[t]
\centerline{\includegraphics[width=0.48\textwidth]{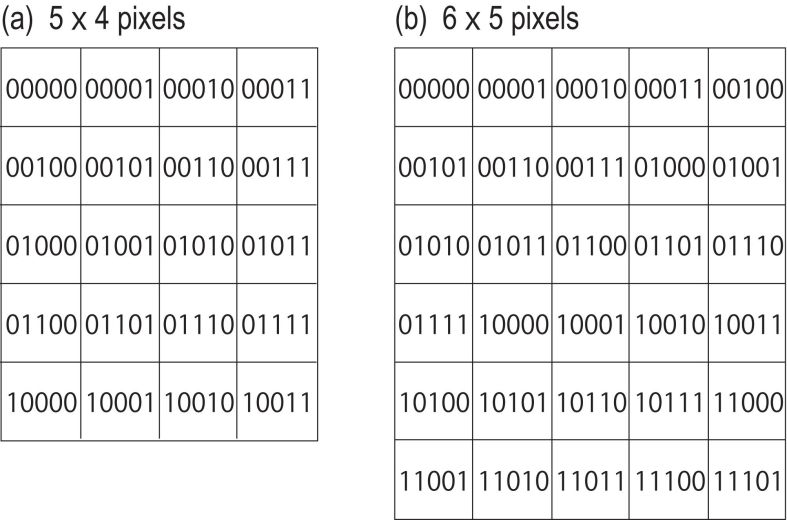}}
\caption{Assignment of binary numbers to (a) $5\times 4$ pixels for the
number recognition and (b) $6\times 5$ pixels\ for the alphabet recognition.}
\label{FigPixel}
\end{figure}

A simplest artificial neuron is given by the perceptron model\cite%
{RosenF,McC}. Here, the input and the weight wave functions are given by Eq.(%
\ref{WaveA}) with $x_{j}=\pm 1$ and $w_{j}=\pm 1$. Such states are called
real equally weighted (REW) states. Furthermore, the step function is used
as the activation function,%
\begin{equation}
y=\Theta (\sum_{j}w_{j}x_{j}-h),  \label{PercepAct}
\end{equation}%
where $\Theta $ is a step function with the threshold $h$, $\Theta \left(
x-h\right) =1$ for $x\geq h$ and $\Theta \left( x-h\right) =-1$ for $x<h$.

In our application of artificial neuron to pattern recognition we use REW
states as in the perceptron model but without employing the activation
function (\ref{PercepAct}). We use the inner product itself as the output.

We now discuss how to construct a REW state from the initial state $%
\left\vert 0\right\rangle \!\rangle $, or how to determine $U_{x}$ and $%
U_{w}^{\dag }$ in Eq.(\ref{WaveB}) in the standard quantum-circuit
implementation\cite{Tacc} and also in the electric-circuit implementation.

In the first step, we prepare the equal-coefficient state defined by%
\begin{equation}
\left\vert \psi _{0}\right\rangle =\frac{1}{\sqrt{2^{N}}}%
\sum_{j=0}^{2^{N}-1}\left\vert j\right\rangle \!\rangle \equiv \frac{1}{%
\sqrt{2^{N}}}\sum_{n_{j}=0,1}|n_{1}n_{2}\cdots n_{N}\rangle .  \label{ESS}
\end{equation}%
This is done by way of the Walsh-Hadamard transform of the initial state $%
\left\vert 0\right\rangle \!\rangle $, 
\begin{equation}
\left\vert \psi _{0}\right\rangle =\bigotimes_{s=1}^{N}U_{\text{H}}^{\left(
s\right) }\left\vert 0\right\rangle \!\rangle ,  \label{EqF}
\end{equation}%
where $U_{\text{H}}^{\left( s\right) }$ is the Hadamard gate acting on the $%
s $th qubit.

\begin{figure}[t]
\centerline{\includegraphics[width=0.48\textwidth]{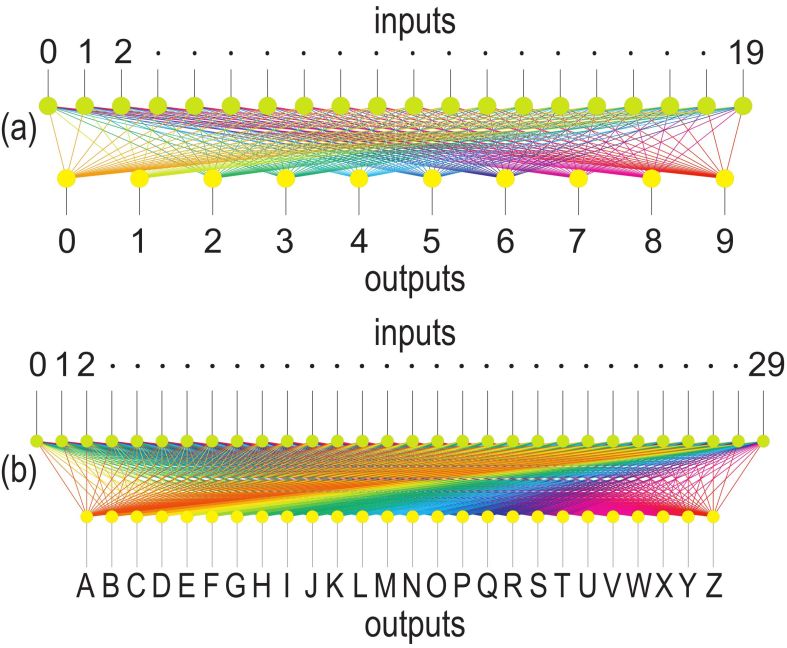}}
\caption{Neural networks for (a) number recognition and (b) alphabet
recognition.}
\label{FigNeuralNet}
\end{figure}

In the second step, we construct $\left\vert \psi _{x}\right\rangle $ and $%
\left\vert \psi _{w}\right\rangle $ from the equal-coefficient state as%
\begin{equation}
\left\vert \psi _{x}\right\rangle =V_{x}\left\vert \psi _{0}\right\rangle
,\qquad \left\vert \psi _{w}\right\rangle =V_{w}^{\dag }\left\vert \psi
_{0}\right\rangle .  \label{EqG}
\end{equation}%
Here, $V_{x}$ is an operation by changing the coefficient $x_{j}$ in the
state $\left\vert \psi _{x}\right\rangle $\ to $-x_{j}$ if $x_{j}=-1$ for
all $j$. Hence, $V_{x}$ is given by a sequential application of C$^{p-1}$Z
gates. For this purpose, we search for the qubit state $\left\vert
j\right\rangle \!\rangle $ whose coefficient is $x_{j}=-1$. Then, we apply
an appropriate C$^{p-1}$Z gate to the state to change its coefficient to $%
x_{j}=1$. An explicit example is given in Appendix A.

We find from (\ref{WaveB}), (\ref{EqF}) and (\ref{EqG}) that%
\begin{equation}
U_{x}=V_{x}\bigotimes_{s=1}^{N}U_{\text{H}}^{\left( s\right) },\quad
U_{w}^{\dag }=V_{w}^{\dag }\bigotimes_{s=1}^{N}U_{\text{H}}^{\left( s\right)
},  \label{EqB}
\end{equation}%
and from (\ref{InnerProd}) and (\ref{EqB}) that%
\begin{equation}
\sum_{j}w_{j}x_{j}=2^{N}\langle \psi _{w}|\psi _{x}\rangle =2^{N}\langle
\!\langle 0|\bigotimes_{s=1}^{N}U_{\text{H}}^{\left( s\right)
}V_{w}V_{x}\bigotimes_{s=1}^{N}U_{\text{H}}^{\left( s\right) }\left\vert
0\right\rangle \!\rangle .  \label{BasicFormula}
\end{equation}%
This is the basic formula to calculate the inner product by a quantum
computer starting from the initial state $\left\vert 0\right\rangle
\!\rangle $. An explicit example of implementation is given in Sec.\ref%
{SecEC}

\begin{figure}[t]
\centerline{\includegraphics[width=0.48\textwidth]{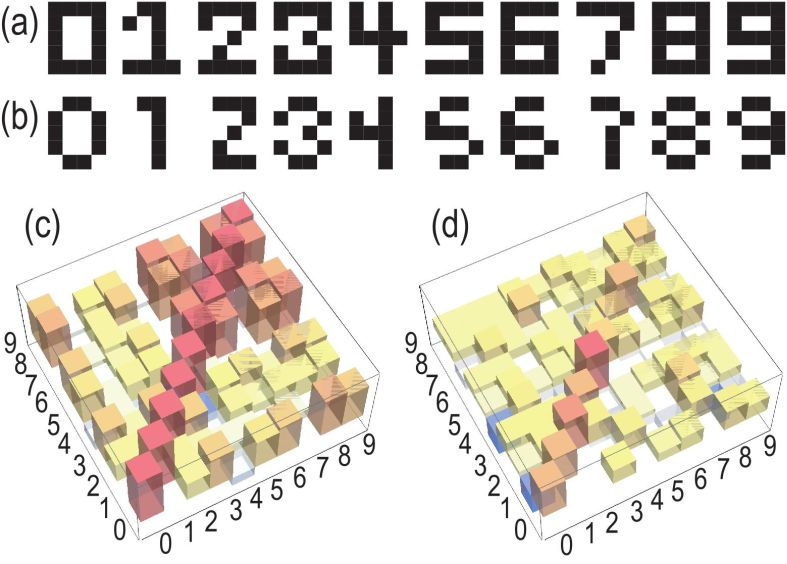}}
\caption{(a) Reference patterns and (b) input patterns of numbers. (c) Self
similarity $\langle \protect\psi _{w}|\protect\psi _{w}\rangle $ and (d)
cross similarity $\langle \protect\psi _{w}|\protect\psi _{x}\rangle $ of
the number recognition. }
\label{FigNumber}
\end{figure}

\section{Pattern recognition}

\label{SecPattern}

Pattern recognition is one of the most useful applications of artificial
neurons. As an example, we consider a pattern made of rectangular pixels
painted in black and white. We show two patterns made of $5\times 4$ pixels
and $6\times 5$ pixels, which are labelled by binary codes as in Fig.\ref%
{FigPixel}(a) and (b). Next, we assign $x_{j}=1$ for white pixel and $%
x_{j}=-1$ for black pixel. Here, $j$ is a decimal number representing a
binary code assigned to a pixel, $0\leq j\leq N_{p}-1$.

In order to represent $N_{x}\times N_{y}$ pixels, we prepare $N$ qubits
satisfying $2^{N-1}<N_{x}\times N_{y}\leq 2^{N}$. These $N$-qubit states are
REW states, which are Eq.(\ref{WaveA}) with $x_{j}=\pm 1$ and $w_{j}=\pm 1$.
Let there be $N_{p}$ patterns to be classified. It is $N_{p}=10$ for the
number recognition and $N_{p}=26$ for the alphabet recognition as in Fig.\ref%
{FigNeuralNet}(a) and (b), respectively. We use a set of reference patterns
as the weight wave function $|\psi _{w}\left( j\right) \rangle $, and
compare them with a set of input patterns $|\psi _{x}\left( j\right) \rangle 
$: Examples are given in Fig.\ref{FigNumber} for $N_{p}=10$ and in Fig.\ref%
{FigAlphaBet} for $N_{p}=26$. In these cases, it is enough to prepare five
qubits. We estimate the similarity between an input pattern and the
reference pattern by calculating the inner product $\langle \psi _{w}|\psi
_{x}\rangle $. We determine which input pattern is most similar to the
reference pattern by searching the largest inner product $\langle \psi
_{w}|\psi _{x}\rangle $. This process is expressed by a single layer neural
network with $N_{x}\times N_{y}$ inputs and $N_{p}$ outputs as in Fig.\ref%
{FigNeuralNet}. The inner product is calculated as%
\begin{equation}
\langle \psi _{w}|\psi _{x}\rangle =\frac{N_{p}-2N_{\text{error}}}{N_{p}},
\end{equation}%
where $N_{\text{error}}$ is the number of errors between the reference and
the input patterns defined by%
\begin{equation}
N_{\text{error}}=\sum_{j=0}^{N_{p}-1}\left( \frac{x_{j}-w_{j}}{2}\right)
^{2}.
\end{equation}%
We note that $\langle \psi _{w}|\psi _{x}\rangle $ can be negative for $N_{%
\text{error}}\geq N_{p}/2$. We find $\left\vert \langle \psi _{w}|\psi
_{x}\rangle \right\vert \leq 1$, where $\langle \psi _{w}|\psi _{x}\rangle
=1 $ indicates the perfect matching.

\begin{figure}[t]
\centerline{\includegraphics[width=0.48\textwidth]{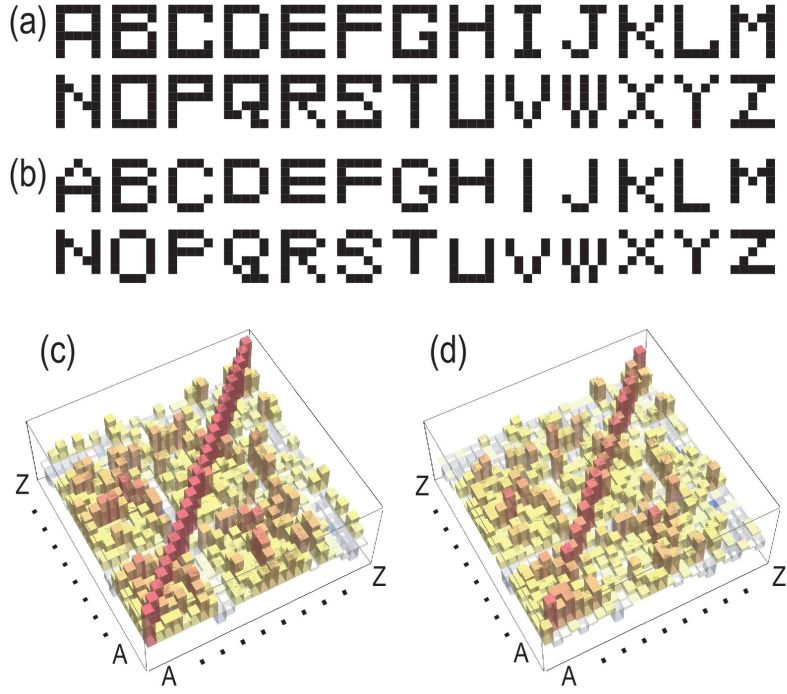}}
\caption{(a) Reference patterns and (b) input patterns of alphabets. (c)
Self similarity $\langle \protect\psi _{w}|\protect\psi _{w}\rangle$ and (d)
cross similarity $\langle \protect\psi _{w}|\protect\psi _{x}\rangle$ of the
alphabet recognition.}
\label{FigAlphaBet}
\end{figure}

As the first example, we study a recognition of numbers. We choose a set of
the reference patterns of numbers as given by Fig.\ref{FigNumber}(a). We
implement them into a wave function $|\psi _{w}\rangle $. Explicit forms are
shown in Appendix B. Then, we take a set of input patterns. See Fig.\ref%
{FigNumber}(b) for an instance. First, we calculate the self similarity
defined by $\langle \psi _{w}\left( j_{1}\right) |\psi _{w}\left(
j_{2}\right) \rangle $, which is shown in Fig.\ref{FigNumber}(c). The
maximum values are taken when $j=j_{1}=j_{2}$ with $\langle \psi _{w}\left(
j\right) |\psi _{w}\left( j\right) \rangle =1$. In order to well recognize
different patterns as different ones, it is necessary that $\langle \psi
_{w}\left( j_{1}\right) |\psi _{w}\left( j_{2}\right) \rangle $ is small for 
$j_{1}\neq j_{2}$. From Fig.\ref{FigNumber}(c), we find that 1, 2, 3, 4, 5
and 7 are well distinguishable because $\langle \psi _{w}\left( j_{1}\right)
|\psi _{w}\left( j_{2}\right) \rangle $ is low. On the other hand, 6, 8 and
9 are hardly distinguishable because the similarity is 0.9, where only one
pixel is different.

Next, we study a cross similarity between the input and the reference
patterns by calculating $\langle \psi _{w}\left( j_{1}\right) |\psi
_{x}\left( j_{2}\right) \rangle $. We fix $j_{2}$ for the input pattern and
determine which reference pattern is most similar by choosing the largest
inner product. We find 0, 1, 2, 3, 4, 5, 6, 7 and 9 are correctly
recognized, but 8 is ill recognized to be 3. See Fig.\ref{FigNumber}(d).

In a similar way, we study an alphabet recognition. We choose a set of the
reference patterns of alphabets as in Fig.\ref{FigAlphaBet}(a) and a set of
input patterns in Fig.\ref{FigAlphaBet}(b). The self-similarity and the
cross-similarity are shown in Fig.\ref{FigAlphaBet}(c) and (d). We find the
following properties from the self similarity: Alphabets are easier to
differentiate comparing to numbers. "F" is hardly differentiated from "P",
where the similarity is 14/15. "C", "D", "G" and "O" are hardly
differentiable among themselves, and "M" and "N" are hardly differentiated
one another, where the similarity is 13/15. We find the following properties
from the cross similarity: There are ill recognitions of "D" to "Q", "E" to
"F", "O" to "D", "X" to "Y" and "Z" to "T". In addition, "C" has equal
similarity to both "C" and "D" in the reference pattern, "G" has equal
similarity to both "D" and "G". "I" has equal similarity to both "I" and
"T". For other cases, the input patterns are well recognized with respect to
the reference patterns.

\section{Complex-artificial neuron}

\label{SecComplex}

We proceed to study a complex-artificial neuron, where the input and the
weight are given by CEW states. Namely, the wave functions are given by (\ref%
{WaveA}) with complex coefficients $x_{j}=e^{i\theta _{j}^{x}}$ and $%
w_{j}=e^{i\theta _{j}^{w}}$. The inner product reads%
\begin{equation}
\langle \psi _{w}|\psi _{x}\rangle =\frac{1}{2^{N}}\sum_{j}w_{j}^{\ast
}x_{j}=\frac{1}{2^{N}}\sum_{j}e^{i\left( \theta _{j}^{x}-\theta
_{j}^{w}\right) }.
\end{equation}%
The output is given by\cite{Mangi}%
\begin{equation}
y=\Phi (\sum_{j}w_{j}^{\ast }x_{j}),
\end{equation}%
where $\Phi $ is a complex-activation function.

\begin{figure}[t]
\centerline{\includegraphics[width=0.43\textwidth]{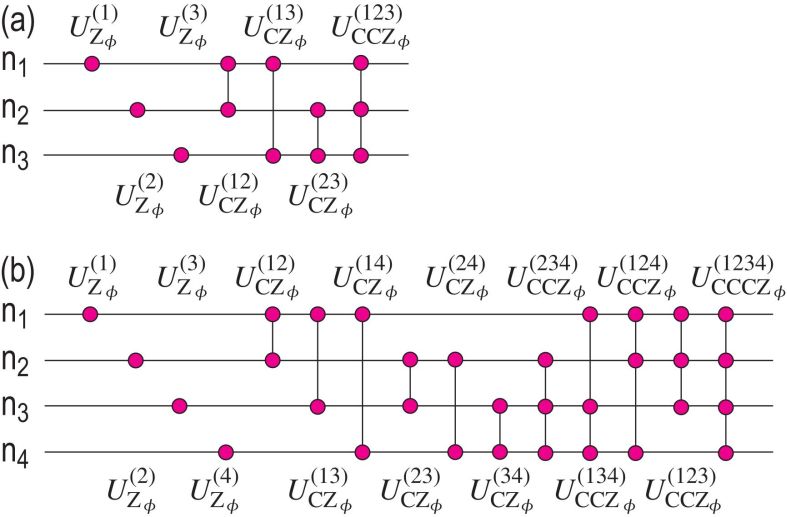}}
\caption{Standard quantum circuits for a generation process of the CEW state
for (a) three qubits and (b) four qubits. An isolated magenta disk indicates
a Z gate. Magenta disks connected by a line indicate a C$^{p-1}$Z$_{\protect%
\phi }$ gate. }
\label{FigWGraph}
\end{figure}

Any CEW state is generated by a sequential application of C$^{p-1}$Z$_{\phi
} $ gates to the equal-coefficient state (\ref{ESS}) precisely as the REW
state is generated by a sequential application of C$^{p-1}$Z gates to it.
Let us explain it by taking the most general CEW state $\left\vert \psi
\right\rangle $ in the $3$-qubit system. It is given by%
\begin{equation}
\left\vert \psi \right\rangle =\sum_{n_{j}=0,1}\alpha
_{n_{1}n_{2}n_{3}}|n_{1}n_{2}n_{3}\rangle ,  \label{EqD}
\end{equation}%
with%
\begin{equation}
\alpha _{n_{1}n_{2}n_{3}}=\frac{1}{\sqrt{2^{N}}}\exp (i\theta
_{n_{1}n_{2}n_{3}}),
\end{equation}%
where we set $\theta _{000}=0$ without loss of generality.

We list up all possible C$^{p-1}$Z$_{\phi }$ gates in Fig.\ref{FigWGraph}%
(a). Recall that all C$^{p-1}$Z$_{\phi }$ gates are commutative. The
generated CEW state is given by%
\begin{align}
& U_{\text{CCZ}_{\phi _{123}}}^{\left( 123\right) }U_{\text{CZ}_{\phi
_{23}}}^{\left( 23\right) }U_{\text{CZ}_{\phi _{13}}}^{\left( 13\right) }U_{%
\text{CZ}_{\phi _{12}}}^{\left( 12\right) }U_{\text{Z}_{\phi _{3}}}^{\left(
3\right) }U_{\text{Z}_{\phi _{2}}}^{\left( 2\right) }U_{\text{Z}_{\phi
_{1}}}^{\left( 1\right) }\bigotimes_{s=1}^{4}U_{\text{H}}^{\left( s\right)
}\left\vert 0\right\rangle \!\rangle  \notag \\
=& \frac{1}{\sqrt{8}}(|000\rangle +e^{i\phi _{3}}|001\rangle +e^{i\phi
_{2}}|010\rangle +e^{i\left( \phi _{2}+\phi _{3}+\phi _{23}\right)
}|011\rangle  \notag \\
& +e^{i\phi _{1}}|100\rangle +e^{i\left( \phi _{1}+\phi _{3}+\phi
_{13}\right) }|101\rangle +e^{i\left( \phi _{1}+\phi _{2}+\phi _{12}\right)
}|110\rangle  \notag \\
& +e^{i\left( \phi _{1}+\phi _{2}+\phi _{3}+\phi _{12}+\phi _{13}+\phi
_{23}+\phi _{123}\right) }|111\rangle ),  \label{EqE}
\end{align}%
where the angle $\phi _{1}$ is that of the Z$_{\phi _{1}}$ gate, $\phi _{12}$
is that of CZ$_{\phi _{12}}$ and $\phi _{123}$ is that of CCZ$_{\phi _{123}}$%
, and so on.

\begin{figure}[t]
\centerline{\includegraphics[width=0.48\textwidth]{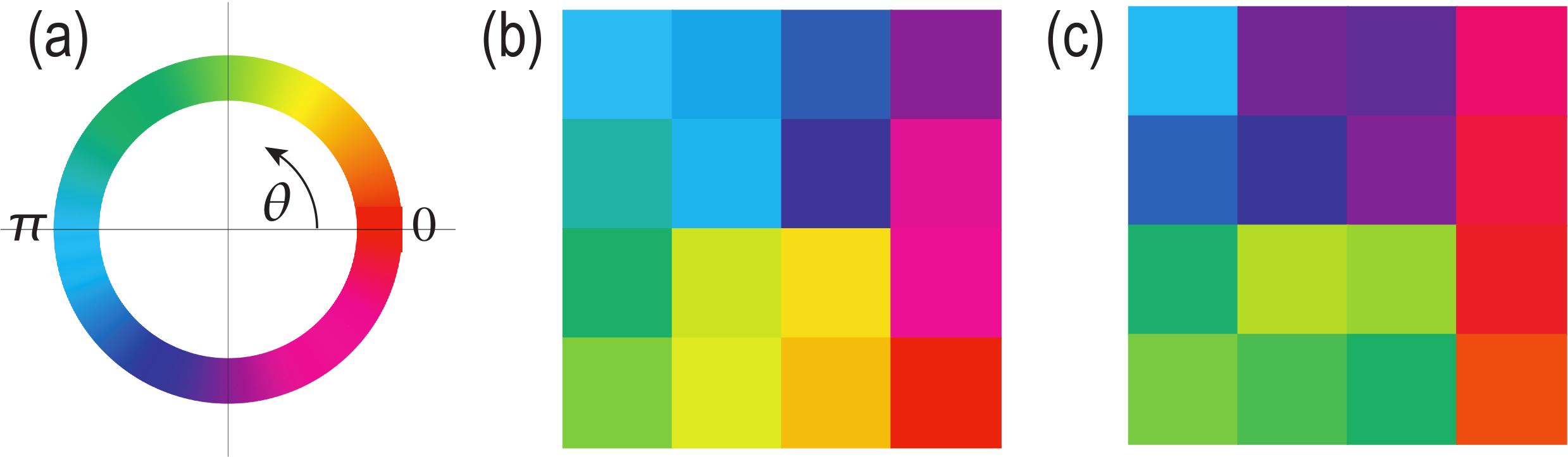}}
\caption{(a) Color circle, (b) reference color pattern and (c) input color
pattern. The input pattern is made by changing randomly colors of the
reference pattern within 20\% randomness.}
\label{FigColor}
\end{figure}

It is easy to see that the angles associated with C$^{p-1}$Z$_{\phi }$ gates
(\ref{EqE}) are uniquely fixed in terms of $\theta _{n_{1}n_{2}n_{3}}$ in
the given CEW state (\ref{EqD}) because there are seven independent
variables in both of these equations. Indeed, by equating (\ref{EqD}) and (%
\ref{EqE}), we obtain relations%
\begin{eqnarray}
\phi _{1} &=&\theta _{100},\quad \phi _{2}=\theta _{010},\quad \phi
_{3}=\theta _{001},  \notag \\
\phi _{12} &=&\theta _{110}-\theta _{100}-\theta _{010},  \notag \\
\phi _{13} &=&\theta _{101}-\theta _{100}-\theta _{001}, \\
\phi _{23} &=&\theta _{011}-\theta _{010}-\theta _{001},  \notag \\
\phi _{123} &=&\theta _{111}+\theta _{100}+\theta _{010}+\theta
_{001}-\theta _{110}-\theta _{101}-\theta _{011}.  \notag
\end{eqnarray}%
We have shown which C$^{p-1}$Z$_{\phi }$ gates we have to prepare in order
to generate the most general CEW state (\ref{EqD}) in the $3$-qubit system.

We also list up all possible C$^{p-1}$Z$_{\phi }$ gates for the $4$-qubit
system in Fig.\ref{FigWGraph}(b). In general, we can always construct an
arbitrary CEW state by applying C$^{p-1}$Z$_{\phi }$ gates in $N$-qubit
systems.

\begin{figure*}[t]
\centerline{\includegraphics[width=0.78\textwidth]{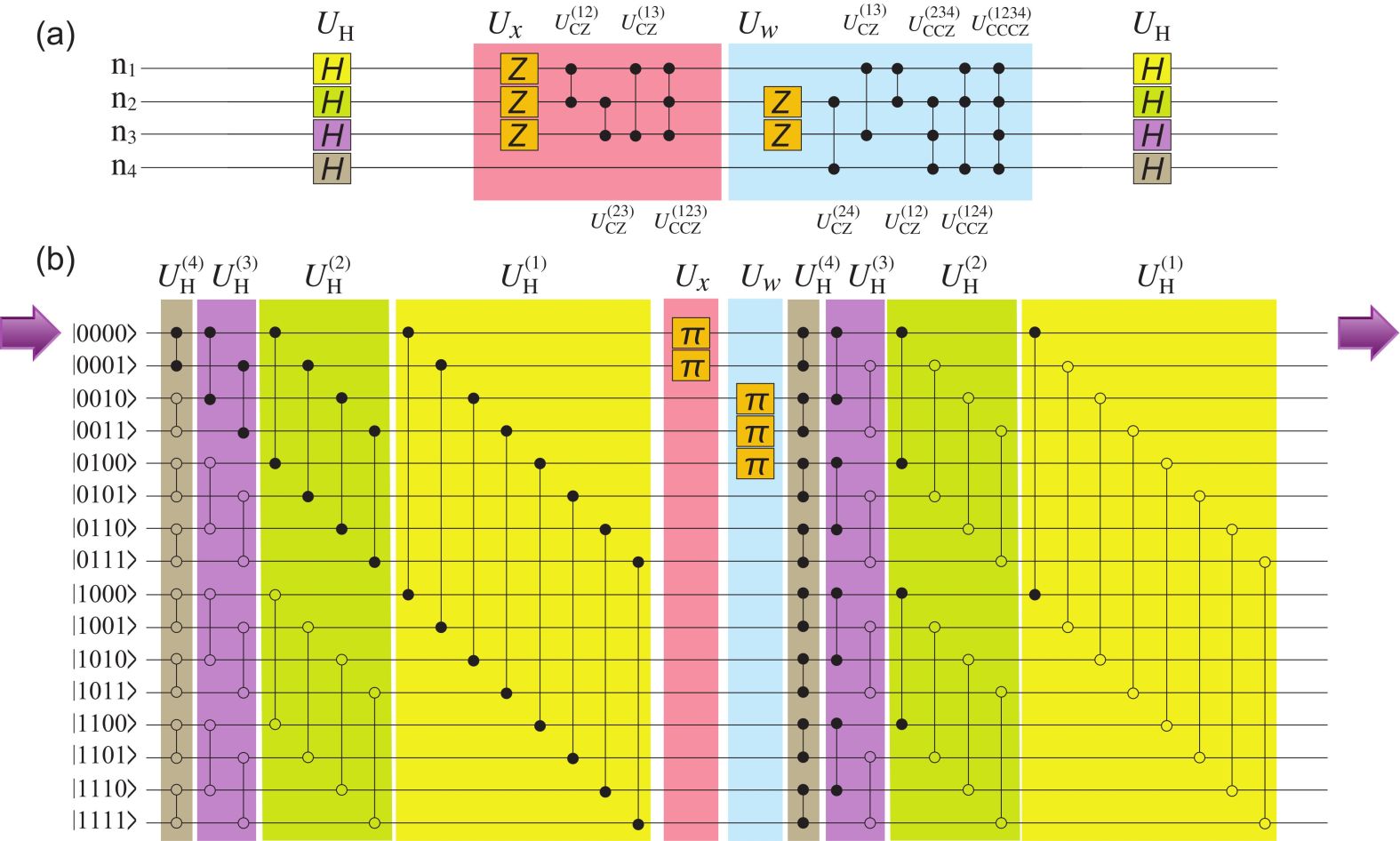}}
\caption{(a) Standard quantum-circuit model for the calculation of the inner
product (\protect\ref{ECBasic}) for an example given by (\protect\ref{StateX}%
) and (\protect\ref{StateW}). We use a four-qubit state $\left\vert
n_{1}n_{2}n_{3}n_{4}\right\rangle $ (b) Corresponding electric-circuit
simulation. The Hadamard gates are denoted by filled or unfilled black disks
connected by a link. The Hadamard bridges with unfilled disks are not
necessary since they are irrelevant for the input state $\left\vert
0\right\rangle \!\rangle $\ and the output state $\langle \!\langle 0|$.}
\label{FigCircuit}
\end{figure*}

\section{Colored pattern recognition}

\label{SecColor}

The color circle is a color pallet indexed by a number on a circle as shown
in Fig.\ref{FigColor}(a). It has a one-to-one correspondence to $e^{i\theta
} $. For example, $\theta =0$ indicates red and $\theta =\pi $ indicates
cyan. Hence, a color pattern made of pixels is well represented by a CEW
state. By using a complex neural network, we can estimate a similarity
between two colored patterns.

For example, we show a reference colored pattern in Fig.\ref{FigColor}(b).
It is enough to prepare 4 qubits for a pattern with 16 pixels. We make an
input colored pattern by modifying color randomly within 20\%. The inner
product of two patterns is $0.993-0.096i$. It is relatively large although
the color of each pixel is modified by 20\%. This is because the input
pattern is created from the reference pattern by adding noise, where the
noise is cancelled by adding all contributions from pixels. Hence, our
scheme can evaluate similarity between two colored patterns with color noise.

The merit of our color representation scheme is that the color circle is
naturally represented by a continuous circle $e^{i\theta }$. In the standard
digital representation, we have to digitalize color. The number of classical
bits increases as the increase of hue decomposition. On the other hand, all
color is continuously represented by one number $e^{i\theta }$ in our scheme.

\section{Electric-circuit implementation}

\label{SecEC}

We implement these models by a set of \textit{LC} resonators. We prepare $%
2^{N}$ \textit{LC} resonators to represent the states $\left\vert
j\right\rangle \!\rangle $ or $\left\vert n_{1}n_{2}\cdots n_{s}\cdots
n_{N}\right\rangle $. The main issue is the electric-circuit implementation
of the inner product formula (\ref{BasicFormula}), or 
\begin{equation}
\langle \psi _{w}|\psi _{x}\rangle =\langle \!\langle
0|\bigotimes_{s=1}^{N}U_{\text{H}}^{\left( s\right)
}V_{w}V_{x}\bigotimes_{s=1}^{N}U_{\text{H}}^{\left( s\right) }\left\vert
0\right\rangle \!\rangle ,  \label{ECBasic}
\end{equation}%
which may be used for CEW states as well as REW states.

The first step is the construction of the equal-coefficient state (\ref{ESS}%
) by applying $\bigotimes_{s=1}^{N}U_{\text{H}}^{\left( s\right) }$ to the
initial state $\left\vert 0\right\rangle \!\rangle $. The action of the
Hadamard transformation $U_{\text{H}}^{\left( s\right) }$ for the $s$th
qubit is simulated by bridging two resonators $\left\vert n_{1}n_{2}\cdots
n_{s}\cdots n_{N}\right\rangle $ and $\left\vert n_{1}n_{2}\cdots \overline{n%
}_{s}\cdots n_{N}\right\rangle $, when $\overline{n}_{s}=1$ for $n_{s}=0$
and $\overline{n}_{s}=0$ for $n_{s}=1$. In the case of $N=4$, the Hadamard
gate $U_{\text{H}}^{\left( 1\right) }$ is simulated by the eight bridges
between%
\begin{equation}
\begin{array}{c}
\left\vert 0000\right\rangle \text{\quad and\quad }\left\vert
1000\right\rangle , \\ 
\left\vert 0001\right\rangle \text{\quad and\quad }\left\vert
1001\right\rangle , \\ 
\left\vert 0010\right\rangle \text{\quad and\quad }\left\vert
1010\right\rangle , \\ 
\left\vert 0011\right\rangle \text{\quad and\quad }\left\vert
1011\right\rangle , \\ 
\left\vert 0100\right\rangle \text{\quad and\quad }\left\vert
1100\right\rangle , \\ 
\left\vert 0101\right\rangle \text{\quad and\quad }\left\vert
1101\right\rangle , \\ 
\left\vert 0110\right\rangle \text{\quad and\quad }\left\vert
1110\right\rangle , \\ 
\left\vert 0111\right\rangle \text{\quad and\quad }\left\vert
1111\right\rangle .%
\end{array}%
\end{equation}%
Although there are many bridges, their assignment is systematic. Apparently
we need $N2^{N-1}$\ operations. Actually, many bridges shown by unfilled
disks in Fig.\ref{FigCircuit}(b) are not necessary since we start with $%
\left\vert 0\right\rangle \!\rangle $\ and end up with $\langle \!\langle 0|$
as in Eq.(\ref{ECBasic}). Then, we may delete all operations which is
irrelevant to the input and the output, which greatly reduces the number of
operations. The necessary operation relating to the $N$-qubit Hadamard gate
is $\sum_{s}^{N}2^{s-1}=2^{N}-1$.The reduction rate is 
\begin{equation}
\lim_{N\rightarrow \infty }\frac{2^{N}-1}{N2^{N-1}}=\lim_{N\rightarrow
\infty }\frac{2}{N}.
\end{equation}

The second step is the operation of C$^{p-1}$Z gates in the case of REW
states. In construct to the application of C$^{p-1}$Z gates in the standard
quantum-circuit implementation, it is enough to apply the $\pi $ phase-shift
only for $\left\vert j\right\rangle \!\rangle $ with $x_{j}=-1$. More
explicitly, the CZ gate for two qubits is simulated by the $\pi $
phase-shift gate applied to the resonator representing $\left\vert
11\right\rangle $, while the CCZ gate for three qubits is simulated by the $%
\pi $ phase-shift gate applied to the resonator representing $\left\vert
111\right\rangle $. In general, the C$^{p-1}$Z gate for $p$ qubits is
simulated by the $\pi $ phase-shift gate applied only to the resonator
representing $\left\vert 11\cdots 1\right\rangle $.

We consider an example of an inner product for the input and the weight
states given by%
\begin{align}
4\left\vert \psi _{x}\right\rangle =& -\left\vert 0000\right\rangle
-\left\vert 0001\right\rangle +\left\vert 0010\right\rangle +\left\vert
0011\right\rangle  \notag \\
& +\left\vert 0100\right\rangle +\left\vert 0101\right\rangle +\left\vert
0110\right\rangle +\left\vert 0111\right\rangle  \notag \\
& +\left\vert 1000\right\rangle +\left\vert 1001\right\rangle +\left\vert
1010\right\rangle +\left\vert 1011\right\rangle  \notag \\
& +\left\vert 1100\right\rangle +\left\vert 1101\right\rangle +\left\vert
1110\right\rangle +\left\vert 1111\right\rangle ,  \label{StateX}
\end{align}%
and%
\begin{align}
4\left\vert \psi _{w}\right\rangle =& \left\vert 0000\right\rangle
+\left\vert 0001\right\rangle -\left\vert 0010\right\rangle -\left\vert
0011\right\rangle  \notag \\
& -\left\vert 0100\right\rangle +\left\vert 0101\right\rangle +\left\vert
0110\right\rangle +\left\vert 0111\right\rangle  \notag \\
& +\left\vert 1000\right\rangle +\left\vert 1001\right\rangle +\left\vert
1010\right\rangle +\left\vert 1011\right\rangle  \notag \\
& +\left\vert 1100\right\rangle +\left\vert 1101\right\rangle +\left\vert
1110\right\rangle +\left\vert 1111\right\rangle ,  \label{StateW}
\end{align}%
where the inner product is $\langle \psi _{w}|\psi _{x}\rangle =3/8$. The
quantum circuit and the electric circuit to calculate\ this inner product
based on the inner product formula (\ref{BasicFormula}) are given in Fig.\ref%
{FigCircuit}(a). Then, we apply the $\pi $ phase-shift gate for $\left\vert
0000\right\rangle $ and $\left\vert 0001\right\rangle $ in order to
construct $|\psi _{x}\rangle $, while we apply the $\pi $ phase-shift gate
for $\left\vert 0010\right\rangle $, $\left\vert 0011\right\rangle $ and $%
\left\vert 0100\right\rangle $ in order to construct $|\psi _{w}\rangle $ as
in Fig.\ref{FigCircuit}(b).

\begin{figure}[t]
\centerline{\includegraphics[width=0.4\textwidth]{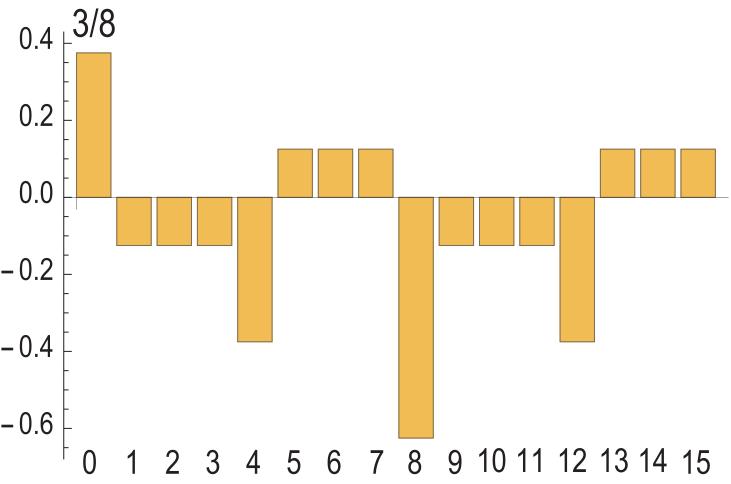}}
\caption{Output $y_{j}$ for each resonator representing $\left\vert
j\right\rangle \!\rangle $. We measure the output $y_0$ for the state $%
\left\vert 0\right\rangle \!\rangle $, which gives an inner product $%
3/8=0.375$.}
\label{FigChart}
\end{figure}

It is convenient to define the output $y_{j}$ for each resonator
representing $\left\vert j\right\rangle \!\rangle $ in Fig.\ref{FigChart} by 
\begin{equation}
\frac{1}{4}\sum_{j=0}^{15}y_{j}\left\vert j\right\rangle \!\rangle =\left(
\bigotimes_{s=1}^{4}U_{\text{H}}^{\left( s\right) }\right) V_{w}V_{x}\left(
\bigotimes_{s=1}^{4}U_{\text{H}}^{\left( s\right) }\right) \left\vert
0000\right\rangle .
\end{equation}%
The outs $y_{j}$ are always real. Especially, we are interested in the
output for $\left\vert 0000\right\rangle $, which is y$_{0}=3/8$. It
reproduce a correct the inner product given below Eq.(\ref{StateW}).

The generalization to the calculation of an inner product of CEW states is
straightforward. As a characteristic feature of the present electric-circuit
simulation, it is possible to create a $\phi $ phase-shift gate with an
arbitrary angle $\phi $ just by tuning the capacitance of the relevant 
\textit{LC} resonator according the formula (\ref{PhaseShift}). Hence, any
CEW state is generated by applying C$^{p-1}$Z$_{\phi }$\ gates with the use
of appropriate $\phi $ phase-shift gates.

\begin{figure}[t]
\centerline{\includegraphics[width=0.4\textwidth]{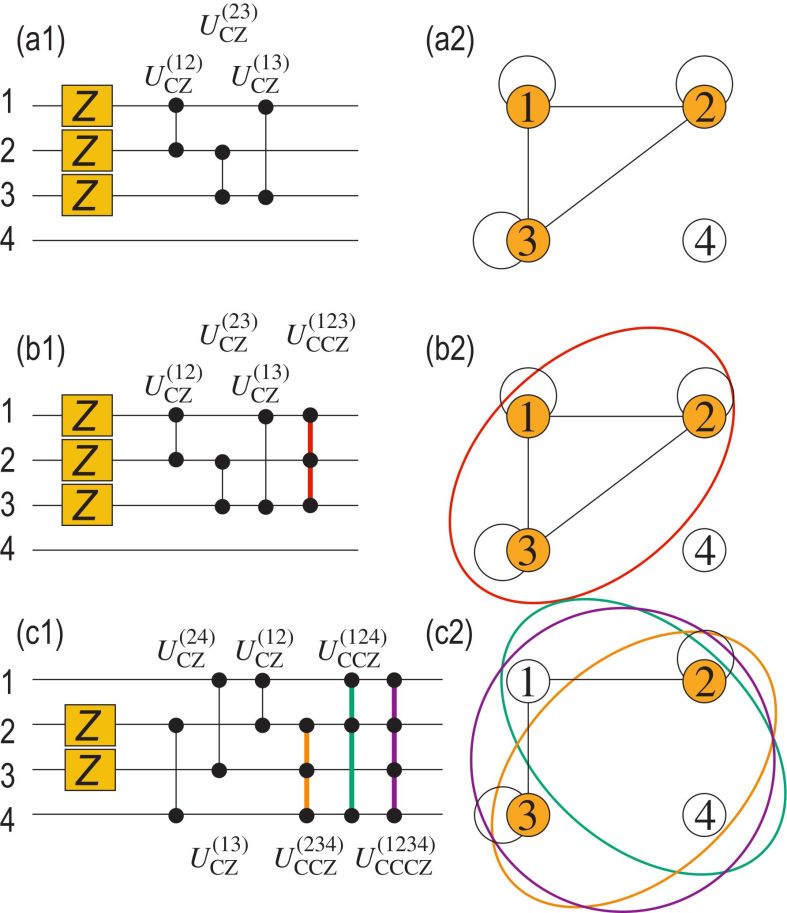}}
\caption{(*1) Quantum-circuit representation, and (*2) graph and hypergraph
representation. Z gates correspond to self-loops, which are marked in orange
disks in a graph or a hypergraph. CZ gates correspond to edges, which are
marked in black lines in a hypergraph. C$^{p-1}$Z gates correspond to
hyperedges, which are marked in colored ovals in a hypergraph. (a*) graph
state, (b*) for $\protect\psi _{x}$ and (c*) for $\protect\psi _{w}$.}
\label{FigHyperGraph}
\end{figure}

\section{Graph theory}

\label{SecGraph}

\textbf{Graph states and hypergraph states.} It is intriguing to examine the
REW state in the context of graph theory. Such a state is referred to as a
graph state\cite{Hein,Hein2,Anders} that is constructed by a sequential
application of Z gates and CZ gates to the equal-coefficient state (\ref{ESS}%
). The order of a Z gate and a CZ gate is irrelevant because they are
diagonal operators and commutable. Then, we may establish one-to-one
correspondence between a graph and a graph state. Indeed, in order to make a
graph corresponding to a graph state, we first prepare $N$ vertices
representing $N$ qubits, as in Fig.\ref{FigHyperGraph}(a2) for an instance
of $N=4$. We add self-loop links to the vertices to which Z gates are
applied, while we connect two vertices by an edge where CZ gates are
operated. Different graphs represent different graph states due to the
commutative nature of the Z and the CZ gates. See Fig.\ref{FigHyperGraph}%
(a2).

The set of all graph states is a subgroup of the REW states by the following
reasoning. The number of the Z gates is $N$, while the number of the CZ
gates is $_{N}C_{2}$. Hence, we can express $2^{N+_{N}C_{2}}$ graph states.
On the other hand, the number of the REW\ states is $2^{2^{N}}$. Here, $%
2^{2^{N}}>2^{N+_{N}C_{2}}$ for $N\geq 3$.

In order to represent a complete set of the REW states, it is necessary to
introduce the notion of hypergraph\cite{HyperGraph,Qu}, which is a
generalization of graph. In a hypergraph, we have a hyperedge connecting
more than three vertices. For example, a CCZ gate is represented by a
hyperedge with order $3$, which connects three\ vertices, as in Fig.\ref%
{FigHyperGraph}(b2). In a similar way, a C$^{p-1}$Z gate is represented by a
hyperedge with order $p$, which connects $p$\ vertices. The number of the C$%
^{p-1}$Z gates is given by $_{N}C_{p}$. Then, the total number of gates is $%
2^{\sum_{p=1}^{N}\left( _{N}C_{p}\right) }=2^{2^{N}-1}$. The overall phase
is irrelevant and thus it is a complete representation. We show a hypergraph
representation of $|\psi _{x}\rangle $ and $|\psi _{w}\rangle $ in Fig.\ref%
{FigHyperGraph}. In Fig.\ref{FigHyperGraph}(b), a hyperedge is represented
by an oval containing three vertices connected by a hyperedge. In Fig.\ref%
{FigHyperGraph}(c), three hyperedges are represented by three ovals
containing vertices connected by three hyperedge.

\textbf{Weighted graph states and hypergraph states.} As a generalization of
graph states and hypergraph states, we may introduce the concepts of
weighted graph states and weighted hypergraph states in the context of CEW
states. A weighted graph state is defined by a sequential application of Z$%
_{\phi }$ gates and CZ$_{\phi }$ gates to the equal-coefficient state (\ref%
{ESS}), where $e^{i\phi }$\ is a weight. Next, a weighted hypergraph state
is generated by a sequential application of C$^{p-1}$Z$_{\phi }$ gates to
the equal-coefficient state (\ref{ESS}). Here, we assign a $p$-hyperedge to
a C$^{p-1}$Z$_{\phi }$ gate as in the case of C$^{p-1}$Z gates, and then we
assign a weight $e^{i\phi }$ to each hyperedge.

\section{Discussion}

\label{SecDisc}

We have proposed an electric-circuit simulation of universal quantum gates
on the basis of \textit{LC} resonators bridged by inductors. Here,
capacitance and inductance are controllable by using a variable capacitance
diode and an active inductor, respectively.

An artificial neuron requires many C$^{p-1}$Z gates for various $p$. It is
actually a hard task to realize C$^{p-1}$Z gates in the standard
quantum-circuit implementation even by employing modern technology such as
superconductor, ion-trap or photonic systems for $p\geq 3$. This difficulty
originates in the fact that a C$^{p-1}$Z gate implies a $p$-body
interaction. Although it is possible to decompose a C$^{p-1}$Z gate into
simpler gates, there emerge many gates\cite{Nielsen}. The problem becomes
worse for a complex-artificial neuron, where we use C$^{p-1}$Z$_{\phi }$
gates instead of C$^{p-1}$Z gates. The C$^{p-1}$Z$_{\phi }$ gate contains
the phase-shift gate with angle $\phi $. It is possible but quite tedious to
construct a C$^{p-1}$Z$_{\phi }$ gate with the use of a set of universal
quantum gates. On the contrary, it is simple to construct a C$^{p-1}$Z$%
_{\phi }$ gate by inserting one $\phi $ phase-shift gate in the
electric-circuit implementation.

Furthermore, it is a nontrivial problem to construct a superposition state
such as REW or CEW states. It is necessary to design several quantum gates
in order to make such a state, for which we need to use a classical computer
in general. See a typical example in Fig.\ref{FigCircuit}(a) and Appendix A.
On the other hand, it is sufficient to insert simply some $\phi $
phase-shift gates in the electric-circuit implementation. Although the
implementation of the Hadamard gates is harder, the assignment is systematic
and trivial. See the corresponding example in Fig.\ref{FigCircuit}(b).

We have previously proposed another kind of electric-circuit simulation of
universal quantum gates, where quantum gates are constructed by bridging
telegrapher wires\cite{EzawaUniv,EzawaDirac}. The number of elements of
electric circuits increases as the increase of the number of quantum gates.
On the other hand, the number of the elements is fixed in the present scheme
irrespective to the number of quantum gates because the operation is
performed in time evolution. Another merit comparing to the wire
construction is that the present scheme is programmable because the gate is
applied temporally, which is contrasted to the wire construction where the
gates are constructed by setting wires.

The author is very much grateful to E. Saito and N. Nagaosa for helpful
discussions on the subject. This work is supported by the Grants-in-Aid for
Scientific Research from MEXT KAKENHI (Grants No. JP17K05490 and No.
JP18H03676). This work is also supported by CREST, JST (JPMJCR16F1 and
JPMJCR20T2).

\appendix
\begin{widetext}

\section{Example of hypergraph generation process}

A REW state $|\psi _{x}\rangle $ is given by Eq.(\ref{WaveA}) with $%
x_{j}=\pm 1$. We explain how to create this state from the
equal-coefficient state (\ref{ESS}). Alternatively, we explain how to
transform this state to the equal-coefficient state.

Let us explicitly study an example given by 
\begin{align}
4|\psi _{x}\rangle =& -\left\vert 0000\right\rangle -\left\vert
0001\right\rangle +\left\vert 0010\right\rangle +\left\vert
0011\right\rangle +\left\vert 0100\right\rangle +\left\vert
0101\right\rangle +\left\vert 0110\right\rangle +\left\vert 0111\right\rangle
\notag \\
& +\left\vert 1000\right\rangle +\left\vert 1001\right\rangle +\left\vert
1010\right\rangle +\left\vert 1011\right\rangle +\left\vert
1100\right\rangle +\left\vert 1101\right\rangle +\left\vert
1110\right\rangle +\left\vert 1111\right\rangle .
\end{align}%
First, we rewrite it so that the coefficient of $\left\vert
0000\right\rangle $ is $1$, 
\begin{align}
4|\psi _{x}\rangle =& -(\left\vert 0000\right\rangle +\left\vert
0001\right\rangle -\left\vert 0010\right\rangle -\left\vert
0011\right\rangle -\left\vert 0100\right\rangle -\left\vert
0101\right\rangle -\left\vert 0110\right\rangle -\left\vert 0111\right\rangle
\notag \\
& -\left\vert 1000\right\rangle -\left\vert 1001\right\rangle -\left\vert
1010\right\rangle -\left\vert 1011\right\rangle -\left\vert
1100\right\rangle -\left\vert 1101\right\rangle -\left\vert
1110\right\rangle -\left\vert 1111\right\rangle ).
\end{align}%
We note that overall phase $-$ is irrelevant in quantum computation.

(i) We focus on the states $|n_{1}n_{2}n_{3}n_{4}\rangle $ such that $%
\sum_{i}n_{i}=1$, among which the coefficients of $\left\vert
1000\right\rangle $, $\left\vert 0100\right\rangle $ and $\left\vert
0010\right\rangle $ are $-1$, while the coefficient of $\left\vert
0001\right\rangle $ is $1$. Then, we apply three $Z$ gates to the first,
second and third qubits, and obtain 
\begin{align}
4U_{\text{Z}}^{\left( 1\right) }U_{\text{Z}}^{\left( 2\right) }U_{\text{Z}%
}^{\left( 3\right) }\psi _{x}=& -(\left\vert 0000\right\rangle +\left\vert
0001\right\rangle +\left\vert 0010\right\rangle +\left\vert
0011\right\rangle +\left\vert 0100\right\rangle +\left\vert
0101\right\rangle -\left\vert 0110\right\rangle -\left\vert 0111\right\rangle
\notag \\
& +\left\vert 1000\right\rangle +\left\vert 1001\right\rangle -\left\vert
1010\right\rangle -\left\vert 1011\right\rangle -\left\vert
1100\right\rangle -\left\vert 1101\right\rangle +\left\vert
1110\right\rangle +\left\vert 1111\right\rangle )
\end{align}

(ii) We focus on the states $|n_{1}n_{2}n_{3}n_{4}\rangle $ such that $%
\sum_{i}n_{i}=2$, among which the coefficients of $\left\vert
0110\right\rangle $, $\left\vert 1010\right\rangle $ and $\left\vert
1100\right\rangle $ are $-1$, while the coefficient of $\left\vert
0011\right\rangle $, $\left\vert 0101\right\rangle $ and $\left\vert
1001\right\rangle $ is $1$. Then, we apply three $CZ$ gates, and obtain 
\begin{align}
4U_{\text{CZ}}^{\left( 13\right) }U_{\text{CZ}}^{\left( 23\right) }U_{\text{CZ%
}}^{\left( 12\right) }U_{\text{Z}}^{\left( 1\right) }U_{\text{Z}}^{\left(
2\right) }U_{\text{Z}}^{\left( 3\right) }\psi _{x}=& -(\left\vert
0000\right\rangle +\left\vert 0001\right\rangle +\left\vert
0010\right\rangle +\left\vert 0011\right\rangle +\left\vert
0100\right\rangle +\left\vert 0101\right\rangle +\left\vert
0110\right\rangle +\left\vert 0111\right\rangle  \notag \\
& +\left\vert 1000\right\rangle +\left\vert 1001\right\rangle +\left\vert
1010\right\rangle +\left\vert 1011\right\rangle +\left\vert
1100\right\rangle +\left\vert 1101\right\rangle -\left\vert
1110\right\rangle -\left\vert 1111\right\rangle )
\end{align}

(iii) We focus on the states $|n_{1}n_{2}n_{3}n_{4}\rangle $ such that $%
\sum_{i}n_{i}=3$, among which the coefficients of $\left\vert
1110\right\rangle $ is $-1$, while the coefficient of $\left\vert
0111\right\rangle $, $\left\vert 1011\right\rangle $ and $\left\vert
1101\right\rangle $ is $1$. Then, we apply one $CCZ$ gate, and obtain the
equal-coefficient state,%
\begin{align}
4U_{\text{CCZ}}^{\left( 123\right) }U_{\text{CZ}}^{\left( 13\right) }U_{%
\text{CZ}}^{\left( 23\right) }U_{\text{CZ}}^{\left( 12\right) }U_{\text{Z}%
}^{\left( 1\right) }U_{\text{Z}}^{\left( 2\right) }U_{\text{Z}}^{\left(
3\right) }\psi _{x}=& -(\left\vert 0000\right\rangle +\left\vert
0001\right\rangle +\left\vert 0010\right\rangle +\left\vert
0011\right\rangle +\left\vert 0100\right\rangle +\left\vert
0101\right\rangle +\left\vert 0110\right\rangle +\left\vert 0111\right\rangle
\notag \\
& +\left\vert 1000\right\rangle +\left\vert 1001\right\rangle +\left\vert
1010\right\rangle +\left\vert 1011\right\rangle +\left\vert
1100\right\rangle +\left\vert 1101\right\rangle +\left\vert
1110\right\rangle +\left\vert 1111\right\rangle ).
\end{align}%
Hence, it follows from (\ref{EqG}) that%
\begin{equation}
V_{x}^{-1}=-U_{\text{CCZ}}^{\left( 123\right) }U_{\text{CZ}}^{\left(
13\right) }U_{\text{CZ}}^{\left( 23\right) }U_{\text{CZ}}^{\left( 12\right)
}U_{\text{Z}}^{\left( 1\right) }U_{\text{Z}}^{\left( 2\right) }U_{\text{Z}%
}^{\left( 3\right) }.
\end{equation}%
Consequently, $|\psi _{x}\rangle $ is obtained by the inverse process as
\begin{equation}
\psi _{x}=-U_{\text{CCZ}}^{\left( 123\right) }U_{\text{CZ}}^{\left(
13\right) }U_{\text{CZ}}^{\left( 23\right) }U_{\text{CZ}}^{\left( 12\right)
}U_{\text{Z}}^{\left( 1\right) }U_{\text{Z}}^{\left( 2\right) }U_{\text{Z}%
}^{\left( 3\right) }\bigotimes_{s=1}^{4}U_{\text{H}}^{\left( s\right)
}\left\vert 0000\right\rangle ,
\end{equation}%
because all C$^{p-1}$Z gates are commutative and one C$^{p-1}$Z gate $U$
satisfies $U^{2}=1$.

In a similar way, 
\begin{align}
\psi _{w}=& \frac{1}{4}(\left\vert 0000\right\rangle +\left\vert 0001\right\rangle
-\left\vert 0010\right\rangle -\left\vert 0011\right\rangle -\left\vert
0100\right\rangle +\left\vert 0101\right\rangle +\left\vert
0110\right\rangle +\left\vert 0111\right\rangle  \notag \\
& +\left\vert 1000\right\rangle +\left\vert 1001\right\rangle +\left\vert
1010\right\rangle +\left\vert 1011\right\rangle +\left\vert
1100\right\rangle +\left\vert 1101\right\rangle +\left\vert
1110\right\rangle +\left\vert 1111\right\rangle )
\end{align}%
is generated as
\begin{equation}
\psi _{w}=U_{\text{CCCZ}}^{\left( 1234\right) }U_{\text{CCZ}}^{\left(
124\right) }U_{\text{CCZ}}^{\left( 234\right) }U_{\text{CZ}}^{\left(
12\right) }U_{\text{CZ}}^{\left( 13\right) }U_{\text{CZ}}^{\left( 24\right)
}U_{\text{Z}}^{\left( 2\right) }U_{\text{Z}}^{\left( 3\right)
}\bigotimes_{s=1}^{4}U_{\text{H}}^{\left( s\right) }\left\vert
0000\right\rangle .
\end{equation}%
Thus, it is a nontrivial problem to construct hypergraph generation circuits
in the standard quantum-circuit implementation.

\section{Number recognition}

We show binary representations of the pattern of each number for the
reference and the input data in Fig.\ref{FigNumber}, where the left hand
side stands for the reference and the right one for the input data,%

\begin{equation}
\begin{array}{c}
0:\left( 11111001100110011111\right) ,\qquad \left(
01101001100110010110\right) , \\ 
1:\left( 01101010001000101111\right) ,\qquad \left(
01100010001000100010\right) , \\ 
2:\left( 11111001001001001111\right) ,\qquad \left(
01110001001001000111\right) , \\ 
3:\left( 01101001001010010110\right) ,\qquad \left(
11111001001010011111\right) , \\ 
4:\left( 10101010111100100010\right) ,\qquad \left(
00101010111000100010\right) , \\ 
5:\left( 11111000111100011111\right) ,\qquad \left(
01101000111000011110\right) , \\ 
6:\left( 11111000111110011111\right) ,\qquad \left(
01101000111010011110\right) , \\ 
7:\left( 11110001001000100100\right) ,\qquad \left(
01100001001000100010\right) , \\ 
8:\left( 11111001111110011111\right) ,\qquad \left(
01101001011010010110\right) , \\ 
9:\left( 11111001111100011111\right) ,\qquad \left(
01101001011100010110\right) ,%
\end{array}%
\end{equation}%
where $0$ indicates a white pixel and $1$ indicates a black pixel.
\end{widetext}

\end{document}